\documentclass[fleqn,usenatbib]{mnras}

\usepackage{newtxtext,newtxmath}

\usepackage[T1]{fontenc}
\usepackage{ae,aecompl}

\usepackage[usenames,dvipsnames]{xcolor}

\newcommand{\sfg}{S$^4$G}
\newcommand{\upperlower}[3]{${#1}^{+{#2}}_{-{#3}}$}
\newcommand{\sparcfire}{\textsc{SpArc\-FiRe}}
\newcommand{\hi}{H\textsc{i}}
\newcommand{\sdsshi}{SDSS+H\textsc{i}}

\usepackage{graphicx}	
\usepackage{amsmath}	
\usepackage{amssymb}	

\title[Galaxy Zoo: constraining the origin of spiral arms]{Galaxy Zoo: constraining the origin of spiral arms}  

\author[Hart et al.]{Ross E. Hart,$^{1}$\thanks{E-mail: ross.hart@nottingham.ac.uk}
Steven P. Bamford,$^{1}$\thanks{E-mail: steven.bamford@nottingham.ac.uk} William C. Keel,$^{2}$ Sandor J. Kruk,$^{3}$ \newauthor Karen L. Masters,$^{4,5}$  Brooke D. Simmons$^{6}$ and Rebecca J. Smethurst$^{1}$\\
$^{1}$School of Physics \& Astronomy, The University of Nottingham, University Park, Nottingham NG7 2RD, UK\\
$^{2}$Department of Physics and Astronomy, University of Alabama, Box 870324, Tuscaloosa, AL 35487, USA\\
$^{3}$Oxford Astrophysics, Denys Wilkinson Building, Keble Road, Oxford OX1 3RH, UK\\
$^{4}$Institute for Cosmology and Gravitation, University of Portsmouth, Dennis Sciama Building, Portsmouth PO1 3FX, UK\\
$^{5}$Department of Physics and Astronomy, Haverford College, 370 Lancaster Ave, Haverford, PA 19041-1392, USA\\
$^{6}$Center for Astrophysics and Space Sciences (CASS), Department of Physics, University of California, San Diego, CA 92093, USA\\
}

\date{Accepted XXX. Received YYY; in original form ZZZ}

\pubyear{2015}

\begin{document}
\label{firstpage}
\pagerange{\pageref{firstpage}--\pageref{lastpage}}
\maketitle

\begin{abstract}
Since the discovery that the majority of low-redshift galaxies exhibit some level of spiral structure, a number of theories have been proposed as to why these patterns exist. A popular explanation is a process known as swing amplification, yet there is no observational evidence to prove that such a mechanism is at play. By using a number of measured properties of galaxies, and scaling relations where there are no direct measurements, we model samples of SDSS and \sfg{} spiral galaxies in terms of their relative halo, bulge and disc mass and size. Using these models, we test predictions of swing amplification theory with respect to directly measured spiral arm numbers from Galaxy Zoo 2. We find that neither a universal cored or cuspy inner dark matter profile can correctly predict observed numbers of arms in galaxies. However, by invoking a halo contraction/expansion model, a clear bimodality in the spiral galaxy population emerges. Approximately 40 per cent of unbarred spiral galaxies at $z \lesssim 0.1$ and $\mathrm{M_*} \gtrsim 10^{10} \mathrm{M_\odot}$ have spiral arms that can be modelled by swing amplification. This population display a significant correlation between predicted and observed spiral arm numbers, evidence that they are swing amplified modes. The remainder are dominated by two-arm systems for which the model predicts significantly higher arm numbers. These are likely driven by tidal interactions or other mechanisms.

\end{abstract}




\section{Introduction}
\label{sec:introduction}

A significant fraction of the local galaxy population display discs with spiral structure, and gaining an understanding why spiral patterns exist has been the subject of numerous studies. A multitude of mechanisms have been proposed to explain the existence of spiral arms. The existence of two arm `grand design' spirals is predicted by density wave theory \citep{Lindblad_63,Lin_64} and tidal interactions \citep{Toomre_72,Tully_74,Oh_08,Elmegreen_83,Dobbs_10}. An alternative hypothesis, known as swing amplification \citep{Goldreich_65b,Julian_66,Goldreich_78,Toomre_81} has been proposed as a mechanism via which most types of spiral arms that are observed in the local Universe, from grand design to flocculent, can be produced. However, a consistent theory to describe all galaxy spiral structure is elusive, and a single mechanism may not be responsible for all types of observed spiral structure.

Swing amplification itself is a manifestation of a balance between shear and self gravity. Self gravity tends to form structures, and shear tends to break up the largest structures over time. Spiral arms form due to unstable regions where self gravity dominates, or from initially leading density waves, but are eventually broken up by the disc shear. In the swing amplified mechanism, leading waves, or regions of density enhancement caused by self gravity, are amplified to stationary, trailing wave patterns around the corotation radius (we refer the reader to Sec.~2.1.3 of \citealt{Dobbs_14} for a more detailed description of swing amplification). Spiral arms can be transient in nature, but a long-lived swing amplified mode can exist in galaxy discs over several rotations \citep{Grand_12,Donghia_13,Sellwood_14}: although spiral arms can be broken and re-made, the average total spiral arm number, or dominant mode, will exist beyond the lifetime of a single spiral arm. The nature of these long-lived modes is directly related to the underlying mass distribution of these galaxy discs. Notably, swing amplified models have predicted that spiral arm number should depend on the underlying mass distribution in spiral galaxies \citep{Athanassoula_87,Athanassoula_88,Bosma_99,Fuchs_99a,Fuchs_99b,Fuchs_04,Fuchs_08}. Spiral arm numbers \citep{Donghia_13,Donghia_15} and pitch angles \citep{Baba_13,Michikoshi_14,Michikoshi_16} can now be predicted directly from the mathematics of swing amplification.

Simulations have the potential to shed some light on what mechanisms are at play in spiral galaxies. Disc simulations give us a unique opportunity to study how forces act to introduce or amplify spiral arms. The earliest N-body simulations struggled to produce realistic spiral arms, with two-arm modes quickly leading to the growth of a bar \citep{Miller_70,Hohl_71,Kalnajs_74,Zang_76}. In order to stop the rapid formation of a bar, \citet{Ostriker_73} (and also \citealt{Hohl_76}) demonstrated that the addition of a spherical dark matter halo component makes cold discs more stable. Early simulations did, however, model galaxies with rigid dark matter haloes; \citet{Athanassoula_02} showed that discs embedded in massive dark matter haloes can still form bars, if a live model with interaction between the halo and the disc is considered. It seems that discs and haloes exist in a somewhat complicated relationship, and both play a role shaping the spiral structure in galaxies. A result of particular interest from the latest simulations of spiral structure is that spiral arm patterns may exist as long-lived modes seeded by small density perturbations in the disc \citep{Fujii_11,Wada_11,Grand_12,Donghia_13}. Such spiral arms arise due to local density perturbations via a swing amplified mechanism.


A key issue for any simulation is directly reproducing observable properties of spiral galaxies. There is still much conflict, with disc simulations usually predicting dominant many-arm modes in galaxy discs. Observations instead suggest that even in unbarred galaxies, two-arm spirals are the most common type of spiral structure \citep{Elmegreen_82,Hart_16} which do not arise as readily in the simulations \citep{Donghia_15}. Therefore, there may be a number of mechanisms responsible for the different spiral arm structures we observe, and all spiral galaxies may not be governed by a dominant swing amplified mode.

The aims of this paper are twofold. We first carefully obtain predictions from swing amplification for samples of real galaxies. These predictions are then compared to observed spiral arm properties, in order to evaluate the performance of the swing amplification model. Swing amplification predicts both the spiral arm number and the pitch angle in galaxies with respect to the relative masses and sizes of the dark matter halo, disc and bulge in galaxies. We combine measurements of bulge and disc masses and sizes with published dark matter halo scaling relations to predict the arm properties of galaxies. We utilise  a large sample of spirals from the SDSS \citep{York_00} and a smaller sample from \sfg{} \citep{Sheth_10} spiral galaxies. Using these data, predicted spiral arm numbers and pitch angles are compared to the same observed quantities. The paper is organised as follows. In Sec.~\ref{sec:data}, we describe all of the sources for the swing amplified model. These include observables for baryonic masses and sizes, and scaling relations for the dark matter component for which we have no direct measurements. In Sec.~\ref{sec:model}, we describe predictions of arm number and pitch angle for a swing amplified model. These predictions are then tested against their respective observed quantities in Sec.~\ref{sec:results}. In Sec.~\ref{sec:discussion}, the results are discussed in context of the relevant theory and literature. The main conclusions from the analysis are described in Sec.~\ref{sec:conclusions}.  

This paper assumes a flat cosmology with $\Omega_\mathrm{m}=0.3$ and $H_0=70 \mathrm{km s^{-1} \, Mpc^{-1}}$.

\section{Data}
\label{sec:data}

In this paper, the overall characteristics of galaxies are predicted with a swing amplification model, and compared to real visual characteristics in galaxies. The model we employ has three main components -- a galaxy bulge, disc and dark matter halo. Measurements and models for these components are outlined in the rest of this section. 

\subsection{Sample selection and visual morphologies}
\label{sec:sample_selection}

\subsubsection{SDSS}
\label{sec:sdss_sample}

The main sample utilised for this paper is taken from the SDSS main galaxy sample (MGS). The galaxies are taken from the SDSS Data Release 7 (DR7; \citealt{Abazijian_09}). The main galaxy sample is an $r$-band selected sample, brighter than $m_r$=17.77. In this paper, we only consider galaxies which have reliable visual classifications from Galaxy Zoo 2 (GZ2;\citealt{Willett_13}) -- this sample has a brighter $r$-band limit, complete to $m_r$=17.0, avoiding galaxies that are too faint to be reliably classified. We also employ an upper redshift limit of $z=0.085$ in accordance with \citet{Willett_15} and \citet{Hart_16}, a general limit to which classifications remain reliable. In this paper, we are only concerned with how the relative sizes and masses of components affect the overall galaxy spiral arm morphology. For this reason, we make no completeness cuts to the sample, selecting all galaxies in the redshift range $0.02<z \leq 0.085$ brighter than $m_r$=17.0.

Galaxy morphological data are obtained from GZ2. We use the debiased statistics from \citet{Hart_16}\footnote{GZ2 morphological measurements are available at data.galaxyzoo.org} to ensure our results are free of resolution-dependent redshift bias. GZ2 users were presented with optical $gri$ composite images and asked a number of questions, regarding the presence of spiral arms and bars. We apply a cut of $p_\mathrm{spiral} \geq 0.5$ to select a reliable sample of spiral galaxies (see \citealt{Hart_16} for examples of spiral galaxies selected in this way). An inclination cut of $(b/a)_g > 0.4$ is also used to ensure we only select face-on spirals with reliable spiral morphology estimates, the same cut we used in \citet{Hart_17a}. The principal concern of this paper is spiral structure, without the influence of bars, so we also define a clean, unbarred sample of galaxies, with $p_\mathrm{bar} \leq 0.2$. This cut has been used in GZ2 papers before to select unbarred galaxies \citep{Galloway_15,Kruk_18}. 

Spiral arm numbers are obtained from the GZ2 catalogue, depending on the fractional responses to the `how many spiral arms are there?' question. We make use of two arm number statistics in this paper. The first is $m$, the response which had the greatest debiased vote fraction -- this can take the values `1', `2', `3', `4' and `5+'. The second is $m_\mathrm{avg}$, the average arm number from the classifications, described in \citet{Hart_17b}. This can take any value between 1 and 5, where $m_\mathrm{avg}=1$ means all classifiers said a galaxy had one spiral arm, and $m_\mathrm{avg}=5$ means all classifiers said a galaxy had five or more spiral arms. 

Given the lack of directly measured pitch angles in GZ2, we use an automated method to detect spiral arms in galaxies and measure their pitch angles, $\psi$. The code \sparcfire{}\footnote{http://sparcfire.ics.uci.edu/} is used for this purpose. This code automatically detects and fits logarithmic spiral arms to input galaxy images, outputting a number of statistics for each arc. We ran \sparcfire{} on the $r$-band images of galaxies, and reliable arcs were detected as described in \citet{Hart_17b}. Any galaxies which had no reliable spiral arms detected are removed from further analysis of spiral arm pitch angle. We define the spiral arm pitch angle for each galaxy as the arc-length weighted average pitch angle, $\psi_\mathrm{avg}$. Further details can be found in \citet{Hart_17b}. 

\subsubsection{S4G}
\label{sec:s4g_sample}

Our analysis is primarily concerned with testing SDSS galaxies with associated morphological information from Galaxy Zoo and \sparcfire{}. However, as a check of both these data sets and our implementation of the swing amplification model, we additionally compare to an independent dataset. We therefore also include a sample of spiral galaxies from the \sfg{} sample \citep{Sheth_10,Munoz-Mateos_13,Querejeta_15}. This is a low-redshift, volume-limited sample of galaxies closer than $d=40$ Mpc, galactic latitude $\rvert b \rvert > 30$\textdegree{}, brighter than $m_B=15.5$ and larger than $D_{25}=1$ arcmin. H\textsc{i} 21cm line measurements are required for accurate distance determination, so the \sfg{} sample therefore consists of late-type galaxies by design. Unlike the SDSS sample, this sample is observed in the near infra-red, specifically the Spitzer 3.6$\mu$m and 4.5$\mu$m bands. The visual morphologies are from the classifications of \citet{Buta_15}. We select SA galaxies from the \sfg{} database, a sample of 101 galaxies in total. The spiral arm structure is also listed in this catalogue, with galaxies listed as either grand design (G), many-arm (M) or flocculent (F). Spiral arm pitch angles are obtained from \citet{Herrera-Endoqui_15}. All of the galaxies in \sfg{} were visually inspected, and logarithmic spiral arms were drawn and fit to the galaxies. Given that we expect all features to be real spiral arms in these galaxies, the galaxy pitch angle is given by the mean pitch angle of all of the measured spiral arms in each galaxy.

\subsection{Baryonic masses and sizes}
\label{sec:masses_and_sizes}

Galaxy stellar masses and sizes for the SDSS sample are obtained from the photometric decompositions of \citet{Simard_11} and \citet{Mendel_14}. \citet{Simard_11} fitted two-component models in the $g$ and $r$ bands for all SDSS galaxies with \textsc{GIM2D} \citep{Simard_02}. Sizes of the relative bulge and disc components are taken from the \citet{Simard_11} fits to the $r$-band of galaxies. \citet{Simard_11} provides measurements of scale length for the disc and half light radius for the bulge. For our Hernquist bulge, scale lengths are measured by dividing the half light radius by a factor of $1+\sqrt{2}$, as described by Eq.~4 of \citet{Hernquist_90}.

We note that the $r$-band does not directly trace the overall stellar mass, with light dominated by younger stars. We therefore correct the sizes of the bulge and disc components by dividing by a factor of $1.5 \pm 0.2$, given that the near infra-red is usually $\sim 1.5$ times smaller than the optical component in galaxies \citep{Vulcani_14,Kennedy_16}. \citet{Mendel_14} took the fitting a stage further and scaled the component fluxes to match the SDSS $ugriz$ bands and fit spectral energy distributions (SEDs) to both the bulge and disc. The \citet{Mendel_14} catalogue therefore gives an estimate of the the total stellar mass content of the SDSS galaxies in the bulge and disc components. To avoid any spuriously fit galaxies, only galaxies where \citet{Mendel_14} deemed the fit to be either a disc system (type=2) or a bulge+disc system (type=3) were included in any samples used later in this paper. Using the bulge+disc fits assumes that all galaxies have two distinct components, but this is not always the case \citep{Simmons_13,Simmons_17}. With this in mind, for galaxies where the $F$-test statistic for a two-component fit is $\leq 0.32$, the disc-only fit is used, and the bulge mass component is set to 0 -- motivation for this cut is given in App.~\ref{sec:fit_test}.

For the \sfg{} sample, photometric bulge+disc decompositions are again used to determine the masses and sizes of the stellar component of galaxies. Bulge and disc photometry are obtained from the fits to the 3.6$\mu$m band from \citet{Salo_15}. We select only galaxies where either a single disc component or a disc and bulge component are well-fit (quality=5). Given that the near infra-red component follows the underlying stellar mass distributions of galaxies closely, we use the 3.6$\mu$m fits directly, without a scaling like that used for the SDSS sample \citep{Eskew_12,Meidt_12}. In reality, the near infra-red mass to light ratio, $\Upsilon_*$ will also vary with respect to the age of the component considered \citep{McGaugh_16}, with older, redder stellar populations having more mass for a given 3.6$\mu$m luminosity. \citet{Schombert_14} give values of $\Upsilon^{3.6}_* = 0.5$ for discs and $\Upsilon^{3.6}_* = 0.7$ for bulges. Other estimates of disc mass-to-light ratios vary by $\approx 0.1$, and single value $\Upsilon_{*}^{3.6}$ have been shown to be reasonable estimates \citep{Meidt_14}. Given that the bulge contribution is small in the \sfg{} sample (see Sec.~\ref{sec:galaxy_properties}), we assume a constant mass-to-light ratio for all of the components; any variation in $\Upsilon^{3.6}_*$ has little effect on the results.

The fraction of the mass in the bulge and disc component is simply the fraction of the 3.6$\mu$m light in each component from \citet{Salo_15}, and the sizes of each of the components are simply the sizes of the components measured at 3.6$\mu$m.

\subsubsection{HI masses and sizes}
\label{sec:HI_mass_and_size}

For the SDSS sample, a set of \hi{} measurements is also available. This can help address any missing baryonic mass in galaxies, given that a fraction of the mass in galaxy discs is gaseous rather than stellar. For a fraction of the SDSS sample, there are \hi{} masses available from ALFALFA survey measurements of the \hi{} 21cm line. These masses are obtained from the $\alpha$70 data release of the ALFALFA survey \citep{Giovanelli_05,Haynes_11}. Reliable detections are defined as objects with ALFALFA $\mathrm{detcode}=1$ or 2 (described in \citealt{Haynes_11}) and a single SDSS matched optical counterpart in accordance with \citet{Hart_17a}. For the galaxies with no direct measurement, we use \hi{} masses estimated from other galaxy properties. \citet{Teimoorinia_17} fitted an artificial neural network (ANN) to 15 input galaxy parameters to estimate \hi{} masses. These estimates do not rely on a single parameter such as stellar mass or colour, which have been shown to vary systematically with spiral arm number \citep{Hart_17a,Hart_17b}, meaning they should be valid estimates for all galaxies. Different \hi{} estimates have different uncertainties, described by the quantity $C_\mathrm{fgas}$ in \citet{Teimoorinia_17}. We therefore select reliable estimates as galaxies with $C_\mathrm{fgas} \geq 0.5$ and include an uncertainty of 0.22 dex, in accordance with \citet{Teimoorinia_17}. 

\hi{} disc size estimates are obtained from the following scaling relation between \hi{} size and galaxy disc size from \citet{Lelli_16}:\begin{equation}
\log(R_\mathrm{HI}) = (0.86 \pm 0.04) \log(R_\mathrm{d}) + (0.68 \pm 0.03) ,
\label{eq:HI_size}
\end{equation} where $R_\mathrm{HI}$ is the radius at which the \hi{} surface density falls to 1 $\mathrm{M_\odot pc^{-2}}$. It has also been demonstrated that the \hi{} scale length, $r_\mathrm{s,HI}$, is closely related to $R_\mathrm{HI}$ -- we therefore use a further scaling relation to equate the two quantities from \citet{Wang_14}:\begin{equation}
r_\mathrm{s,HI} = (0.19 \pm 0.03) R_\mathrm{HI}.
\label{eq:HI_conversion}
\end{equation}

Using these relations, we create exponential stellar + \hi{} discs. The total disc mass is given by adding the \hi{} mass to the stellar disc mass, and the disc scale length is given by the scale length of the best fitting exponential profile to the stellar plus disc systems. Discs created in this way are later referred to as \sdsshi{} samples.

\subsection{Dark matter haloes}
\label{sec:dark_matter_haloes}

The final component that requires consideration is the dark matter halo, the only component in the model that is not observable. We use published scaling relations between the galaxy dark matter halo mass and galaxy stellar mass to estimate the dark matter halo mass for each galaxy. We use the relation of \citet{Dutton_10}, which combined abundance matching studies and various observational studies of dark matter haloes from satellite kinematics and weak lensing. The best fit line to observational studies from \citet{Mandelbaum_06}, \citet{Conroy_07} and \citet{More_10} for late-type galaxies yielded the following scaling relation:\begin{equation}
\label{eq:halo_mass} 
y = y_0 \Big(\frac{x}{x_0}\Big)^{\alpha} \Big[\frac{1}{2} + \frac{1}{2}\Big(\frac{x}{x_0}\Big)^{\gamma}\Big]^{(\beta - \alpha)/\gamma}. 
\end{equation} The quantity $x$ is the galaxy total stellar mass, $\mathrm{M_{star}}$, and the quantity $y$ is the halo-to-galaxy mass, $y=\mathrm{M_{200}}/\mathrm{M_{star}}$. For late type galaxies, the parameters are $\alpha=-0.50 ^{+0.025}_{-0.075}$, $\beta = 0.0$, $\log(x_0)=10.4$, $\log(y_0)=1.89^{+0.14}_{-0.12}$ and $\gamma=1.0$. We calculate the total halo mass for each of our galaxies using Eq.~\ref{eq:halo_mass} and the total galaxy stellar mass defined in Sec.~\ref{sec:masses_and_sizes}. These are then converted to virial radii, $R_\mathrm{200}$, with (e.g. \citealt{Huang_17}):\begin{equation}
\label{eq:halo_size} 
R_\mathrm{200} = \Big[\frac{3\mathrm{M_{200}}}{4 \pi \cdot 200 \rho_\mathrm{crit}}\Big]^{1/3} , 
\end{equation} where $\rho_\mathrm{crit}$ is the critical density of the Universe at $z=0$. To convert this to a halo scale radius, $a_\mathrm{h}$, we use the relation \begin{equation} 
\label{eq:size_vs_concentration} 
R_{200} = c_{200}a_\mathrm{h} .
\end{equation} In order to measure a scale radius, one requires knowledge of the halo concentration. We again rely on a published scaling relation, this time from N-body simulations which form NFW profiles. The halo concentration is related to the halo mass using the abundance matching equation of \citet{Dutton_14}: \begin{equation} 
\label{eq:concentration} 
\log(c_{200}) = 0.905 - 0.101 \log(\mathrm{M_{200}}/[10^{12}h^{-1}\mathrm{M_\odot}]) . 
\end{equation} From these scaling relations, we compute the total halo mass $\mathrm{M_{200}}$ and the scale length $a_\mathrm{h}$ for each of our galaxies. 

\subsubsection{Halo profiles}
\label{sec:halo_shapes}

\begin{figure}
\centering
\includegraphics[width=0.45\textwidth]{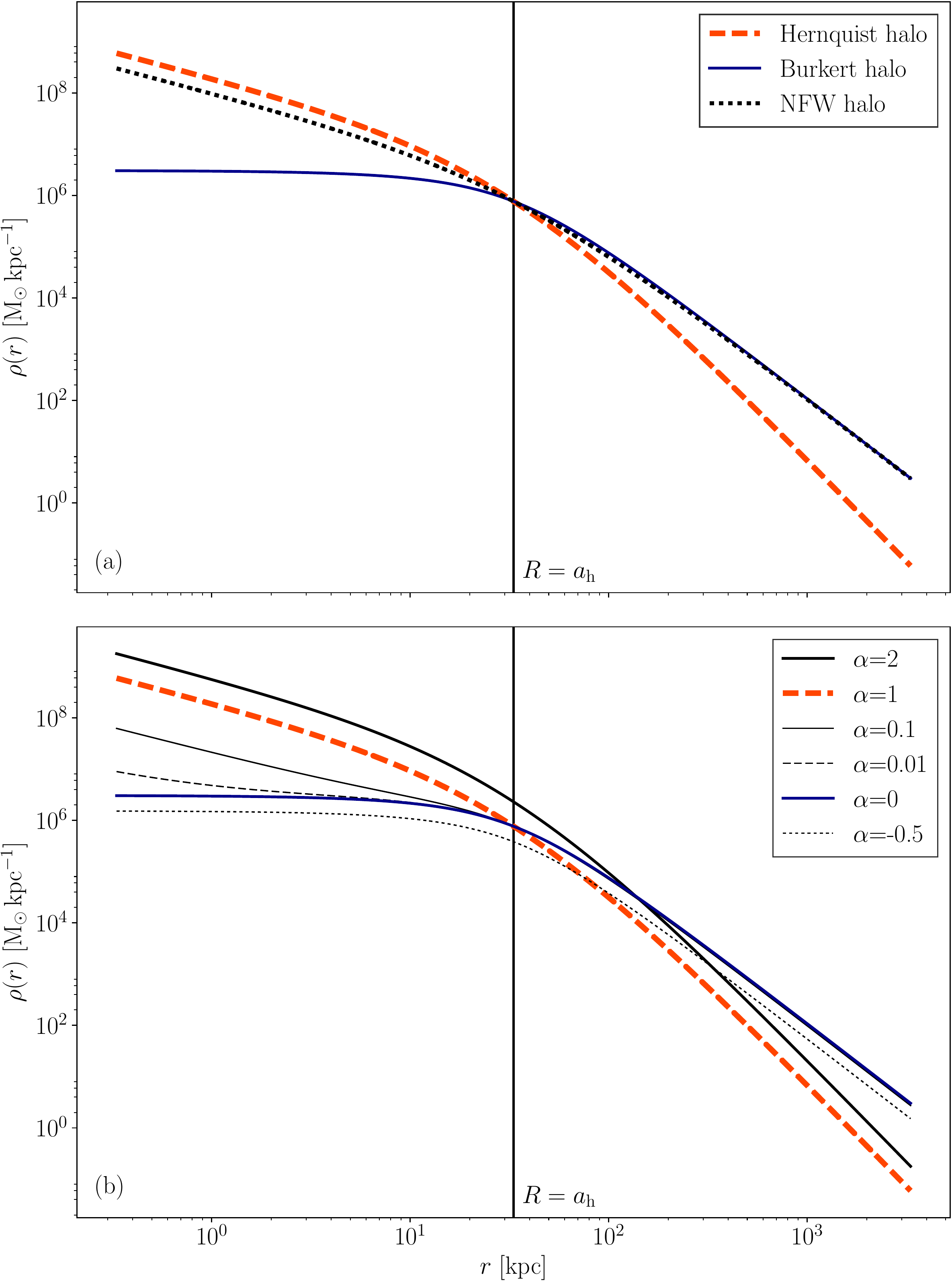}
\caption{(a) Comparison of NFW (black dotted line), Hernquist (orange dashed line) and Burkert (blue line) halo shapes for the Milky Way. The Hernquist halo follows the cuspy shape of the NFW profile, and the Burkert profile is instead cored. (b) Comparison of various values of $\alpha$. When $0 \leq \alpha  \leq 1$, the halo changes between more cored or more cusped. When $\alpha < 0$, the cored shape is retained, but the halo is less massive. When $\alpha > 1$, the cusped shape is retained, but the halo is made more massive.}
\label{fig:halo_shapes}
\end{figure}

For mathematical simplicity, we consider two dark matter profiles in this analysis. The first is the \citet{Hernquist_90} dark matter halo, referred to as `Hernquist' hereafter. This halo has the desirable quality that it closely matches the cusped NFW dark matter haloes \citep*{Navarro_96} in the inner regions. In the outer regions, the dark matter halo begins to deviate from that of the NFW dark matter profile. As it is the inner dark matter profile that is most critical to influencing spiral arm morphology in galaxy discs \citep{Donghia_15}, we choose to match the inner regions closely by matching to the dark matter density at $a_\mathrm{h}$. The shape of the Hernquist halo for a galaxy with parameters from the Milky Way measured in \citet{Bovy_13} is shown by the orange dashed line in Fig.~\ref{fig:halo_shapes}a. 

The Hernquist profile is a classic `cusped' profile predicted by simulations (e.g. \citealt*{Navarro_96}). However, measurements of low surface brightness galaxies reveal that the inner profiles of galaxies are instead more likely to have central `cores' \citep{deBlok_10,Bullock_17}. In order to understand the influence on the shape of the dark matter profile, we adopt a \citet{Burkert_95} dark matter profile for comparison, described as a `Burkert' profile in the rest of this paper. The Burkert profile has a number of characteristics that make it ideal for comparison to the Hernquist profile. It follows a similar shape to the much used NFW profile in the outer regions, which is useful given that the scaling relations we employ in this paper are based upon NFW profiles. However, its centre has a `core' rather than a `cusp', unlike the Hernquist and NFW profiles. In Fig.~\ref{fig:halo_shapes}a, the blue line indicates the Burkert dark matter halo profile for the Milky Way model. Together, these allow us to compare the spiral arm properties of `cusped' and `cored' profiles. The use of the two different dark matter halo profiles also allows us to interpolate between them, a property which is used later in this paper. We define the quantity $\alpha$ to interpolate between a cusped and cored profile. The quantity $\alpha$ is used to give the following dark matter halo profiles: \begin{equation}
\label{eq:alpha}
\begin{aligned}
\rho(r) &= (1+\alpha)\rho_\mathrm{b}(r) &(\alpha < 0) \\
\rho(r) &= (1-\alpha)\rho_\mathrm{b}(r) + \alpha \rho_\mathrm{h}(r) &(0 \leq \alpha \leq 1) \\ 
\rho(r) &= \alpha \rho_\mathrm{h}(r) &(\alpha > 1) , 
\end{aligned}
\end{equation} where $\rho_\mathrm{h} (r)$ and $\rho_\mathrm{b} (r)$ are the densities of the Hernquist and Burkert dark matter profiles at a radius $r$. The influence that the quantity $\alpha$ has on the dark matter halo is shown in Fig.~\ref{fig:halo_shapes}b. A value of $\alpha=1$ means that the dark matter halo is a cusped Hernquist halo and $\alpha=0$ means that the dark matter halo is a cored Burkert halo. Interpolating between the two means that the halo is more or less like the Hernquist and Burkert profiles. To allow for for sensible behaviour outside $0 \leq \alpha \leq 1$, we extrapolate as follows. For values of $\alpha < 0$, the halo shape does not change from that of the Burkert profile, but the total halo mass in the inner regions is reduced. For $\alpha > 1$, the halo stays cusped, but is more massive in the inner regions. 

\subsection{Overall galaxy properties}
\label{sec:galaxy_properties}

\begin{figure*}
\includegraphics[width=0.975\textwidth]{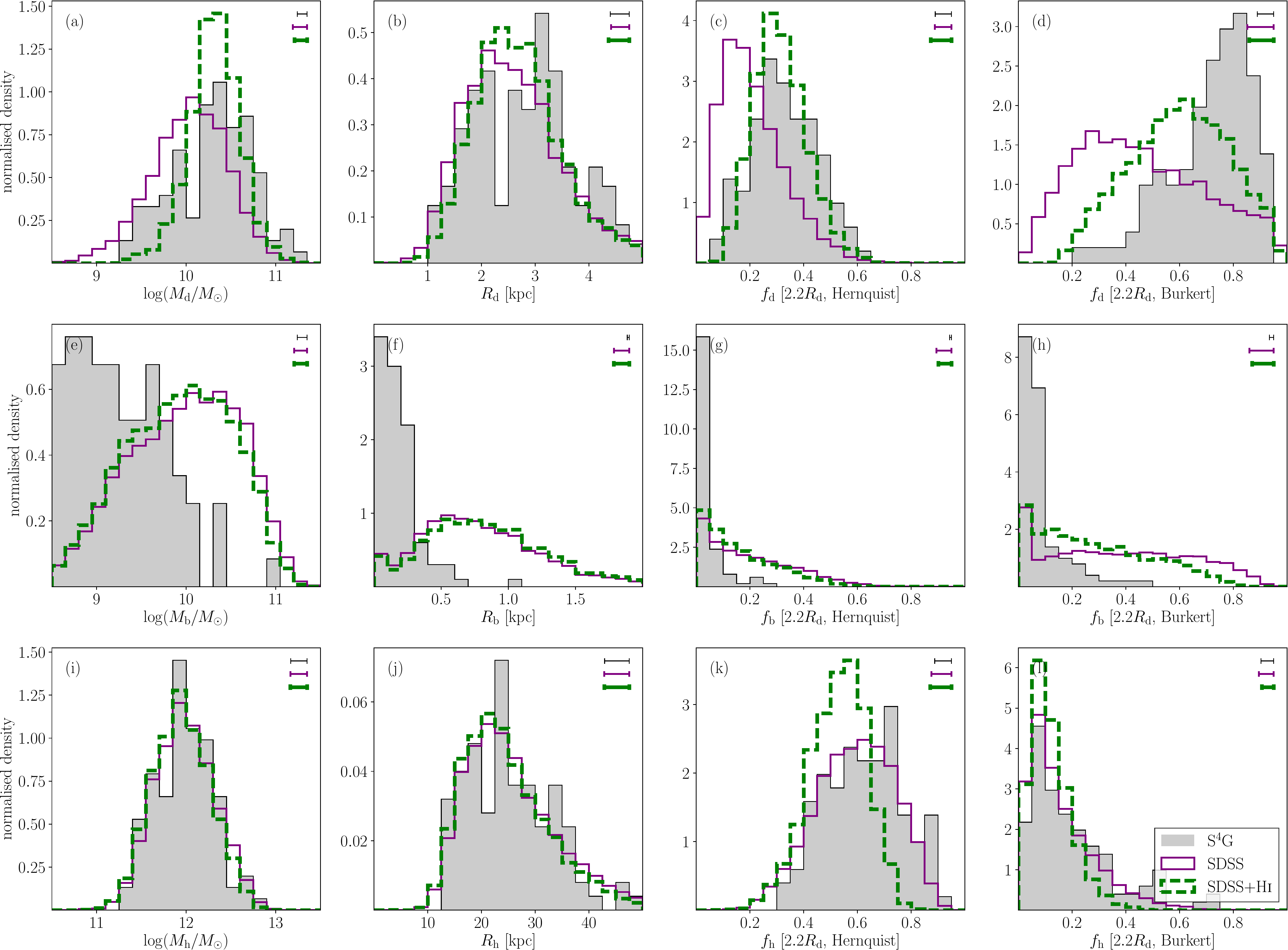}
\caption{Distributions of galaxy stellar mass, scale radius and mass fractions for our galaxy samples. The top row shows the distributions of disc total stellar mass (a), radius (b), fraction of the total galaxy mass within $2.2R_\mathrm{d}$ for the Hernquist halo (c) and the same fraction with a Burkert halo (d). The same four parameters are shown for the galaxy bulge (middle row, e-h) and the galaxy halo (bottom row, i-l). The distributions are shown for the three distributions utilised in this paper (see Sec.~\ref{sec:data}): the \sfg{} sample is shown by the grey filled histograms, the SDSS is shown by the purple stepped histograms and the \sdsshi{} is shown by the thicker dashed green histograms. The median error on the parameters are indicated by the error bars in the upper right of each sub figure.}
\label{fig:all_distributions}
\end{figure*}

Only galaxies with measurements of bulge+disc or disc masses are included in these final samples. The overall numbers of galaxies in each of these samples are listed in the second column of Table.~\ref{table:sample_sizes}. Only some of the galaxies have reliably identified spiral arms from which the pitch angle, $\psi$ can be measured -- the number of galaxies with measured $\psi$ values are shown in the third column of Table.~\ref{table:sample_sizes}. The final column shows the median, 16th and 84th percentiles of the stellar mass of all galaxies in each sample. Despite the different sample selections employed in each of the samples, all of the samples have similar stellar mass distributions with median $\log(\mathrm{M_*})\sim10.4\log(\mathrm{M_\odot})$. 

\begin{table}
\label{table:sample_sizes}
\caption{Number of galaxies in each of the three samples used in this paper. The first column shows the name of each sample. The second column indicates the total number of galaxies in each sample, and the third column indicates how many of these galaxies have measured pitch angles. The final column shows the median, 16th and 84th percentiles of stellar mass for each sample.}

\begin{tabular}{cccc}
\hline
 Sample           &   $N_\mathrm{gal}$ &   $N_\mathrm{gal}$ (with measured $\psi$) & $\log(\mathrm{M/M_\odot})$   \\
\hline
 S$^4$G           &                101 &                                        77 & $10.4 (9.8,10.8)$            \\
 SDSS             &               7611 &                                      2661 & $10.4 (9.9,10.8)$            \\
 \sdsshi{} &               5696 &                                      2241 & $10.3 (9.8,10.7)$            \\
\hline
\end{tabular}

\end{table}

The overall population stellar mass and size characteristics are shown in Fig.~\ref{fig:all_distributions}. The low-redshift galaxies occupy a range of bulge, disc and halo masses. The first column shows the bulge, disc and halo stellar masses for the \sfg{} sample (grey filled histograms), the SDSS sample (purple stepped histograms) and the \sdsshi{} sample (green stepped histograms with dashed lines). The first attribute to note is the change in the disc mass and scale length when the \hi{} is included in the disc fit, shown in Fig.~\ref{fig:all_distributions}a. The median disc mass is \upperlower{10.06}{0.46}{0.39} $\log(\mathrm{M_\odot})$ for the pure stellar disc and increases to \upperlower{10.32}{0.27}{0.26} $\log(\mathrm{M_\odot})$ with the inclusion of H\textsc{i}. The disc radius also increases from \upperlower{2.53}{0.82}{1.07} to \upperlower{2.62}{0.73}{0.90} kpc. These differences lead to differences in the disc fractions, $f_\mathrm{d}$, in Fig.~\ref{fig:all_distributions}c-d, where the inclusion of H$\textsc{i}$ in the discs leads to the discs being more maximal. The disc fraction is defined as the fraction of the total mass inside a given radius that is in the baryonic disc component, $\mathrm{M_{d}}(r)/[\mathrm{M_{b}}(r)+\mathrm{M_{d}}(r)+\mathrm{M_{h}}(r)]$. The inclusion of \hi{} has a strong influence on the disc properties, which may in turn affect the properties of spiral arms, which will be explored in Sec.~\ref{sec:results}.

The next item we note is the clear differences in the bulge properties of galaxies selected for the SDSS and \sfg{} samples. From Fig.~\ref{fig:all_distributions}a-b, we see that the disc properties are consistent in these samples, despite the differing selection criteria. The bulges of the SDSS galaxies have median stellar mass of \upperlower{9.88}{1.07}{0.69} $\log(\mathrm{M_\odot})$ (or \upperlower{9.83}{0.91}{0.67} $\log(\mathrm{M_\odot})$ for the \sdsshi{} sample) and median radius of \upperlower{0.74}{0.38}{0.51} kpc (or \upperlower{0.80}{0.40}{0.50} kpc for the \sdsshi{} sample). However, bulges in the \sfg{} sample are systematically smaller in both mass (Fig.~\ref{fig:all_distributions}e) and size (Fig.~\ref{fig:all_distributions}f), with median values of \upperlower{8.99}{0.64}{0.76} $\log(\mathrm{M_\odot})$ and \upperlower{0.15}{0.07}{0.14} kpc. These offsets could be due to two reasons. The first is sample selection: the SDSS galaxy sample should include all galaxies with spiral morphology, regardless of bulge mass; the selection of late-type galaxies in the \citet{Buta_15} classifications may be slightly different to the ones we employ for our SDSS samples. However, given the careful selection outlined in Sec.~\ref{sec:sample_selection}, we expect both samples to comprise primarily of unbarred, late-type, spirals. The only differences would therefore be caused by the nature of the visual classifications employed. For example, in GZ2, users were not asked to quantify the strength of bar features. The GZ2 sample may therefore comprise some weakly barred galaxies, which would have been detected by the \citet{Buta_15} due to the higher resolution imaging (as the \sfg{} galaxies are selected at a closer distance than the SDSS galaxies, as described in Sec.~\ref{sec:sdss_sample} and \ref{sec:s4g_sample}), and the expert nature of the classifications. The other, likely more significant difference, is that the techniques used to measure bulge mass differ, mainly in the wavelength selected, but also in the image resolution. The SDSS bulge-disc masses are derived from fits to the stellar population of the galaxies using the optical \textit{ugriz} bands. The \sfg{} sample instead uses information from high resolution images of galaxies in the near infra-red, which directly traces the older stellar population and thus the underlying stellar mass distribution. Investigating the true cause of this offset is beyond the scope of this paper, but does highlight the importance of using two complementary datasets to investigate any results. 

\section{The galaxy model}
\label{sec:model}

In this section, we draw upon a number of measured parameters to model spiral galaxies and predict their properties with a swing amplified model.\footnote{The code used to model the galaxies described in this section is publicly available at https://zenodo.org/record/1164581} Wherever there are no directly measurable quantities in galaxies, we use well-defined scaling relations to predict expected properties in galaxies. All of the input quantities to the models defined in this section are described in Sec.~\ref{sec:data}.

\subsection{Swing amplification derived quantities}
\label{sec:swing_amplification_quantities}

We adopt the model described in \citet{Donghia_13} and \citet{Donghia_15} for our spiral galaxies. In \citet{Donghia_15}, an equation was derived from arguments of swing amplification and disc stability, and verified by N-body simulations of isolated discs. The equation describing the dominant spiral arm mode at a given galaxy radius is given by: \begin{equation}
\label{eq:donghia_m}
\begin{split}
m = \frac{e^{2y}}{X} & \Big( \big[\frac{\mathrm{M_b}}{\mathrm{M_d}}\frac{2y+3a_\mathrm{b}/R_\mathrm{d}}{(2y+a_\mathrm{b}/R_\mathrm{d})^3} \big] \\ & + \big[ \frac{\mathrm{M_h}}{\mathrm{M_d}}\frac{2y+3a_\mathrm{h}/R_\mathrm{d}}{(2y+a_\mathrm{h}/R_\mathrm{d})^3} \big] \\ & + y^2(3I_0K_0 - 3I_0K_1 + I_1K_2 - I_2K_1) \\ & + 4y(I_0K_0 - I_1K_1) \Big) . 
\end{split}
\end{equation} The quantity $y=R/2R_\mathrm{d}$, meaning that the predicted spiral arm number can vary with galaxy radius. This equation follows from classic models of swing amplification, first outlined in \citet{Toomre_81}. A useful property of this equation is that it can be split into three main components, each contributing to the expected spiral arm number: the bulge term, the halo term and the disc term. The bulge term, $m_\mathrm{b}$, is given by the first line to the right of the equality in Eq.~\ref{eq:donghia_m}, and depends on the bulge mass ($\mathrm{M_b}$), disc mass ($\mathrm{M_d}$), bulge scale length ($a_\mathrm{b}$) and disc scale length ($R_\mathrm{d}$). The simplicity of this term's form is due to the adoption of a Hernquist profile to model the bulge mass distribution, compared to, for example, a de Vaucoleurs profile \citep{DeV_48}. Generally, galaxies with greater bulge-to-disc mass ratios and galaxies with smaller bulge-to-disc size ratios for a given bulge mass are predicted to have more spiral arms.

The second line to the right of the equality in Eq.~\ref{eq:donghia_m} gives a similar term which we call $m_\mathrm{h}$, this time with the bulge mass and size replaced by halo mass and size ($\mathrm{M_h}$ and $a_\mathrm{h}$). This term is very similar to the one for the the bulge, as \citet{Donghia_15} model the halo with a Hernquist profile. However, there is evidence that galaxy dark matter haloes may be less cuspy than a Hernquist profile (e.g. \citealt{Flores_94,vdB_00,deBlok_01,deBlok_10,Bullock_17}). We therefore derive an alternative form of the halo term for a Burkert dark matter profile in App.~\ref{sec:burkert_profile}. Either way, there is a clear expected dependence on the dark matter halo and disc properties -- galaxies with greater halo-to-disc mass ratios and galaxies with smaller halo-to-disc sizes are predicted to have more spiral arms. 

The final term of the Eq.~\ref{eq:donghia_m} is the disc term, $m_\mathrm{d}$. The mathematical formulation is given in more detail in \citet{Donghia_13}. The quantities $I_0$ and $K_0$ are Bessel functions of the first kind with respect to $y$. From equations \ref{eq:donghia_m} and \ref{eq:burkert_arm_number}, spiral arm numbers can be predicted for the swing amplified model.

Another property we can use to quantify the spiral arms is the pitch angle, $\psi$. The rate of shear has a direct influence on the pitch angle of the spiral arms one expects to measure \citep{Fuchs_01,Seigar_06,Seigar_08,Baba_13}. The shear is given by (e.g. \citealt{Julian_66,Michikoshi_16}): \begin{equation}
\label{eq:Gamma}
\Gamma = 2 - \frac{\kappa^2}{2 \Omega ^2}.
\end{equation} The quantity $\kappa$ is the epicycle frequency, and $\Omega$ is the angular frequency of the system. A falling rotation curve has $\Gamma > 1$, and a rising rotation curve has $\Gamma < 1$. Various conversion factors exist for converting the rate of shear to a pitch angle. Some are based upon observational studies of nearby galaxies \citep{Seigar_06}, others from the analysis of the mathematics of swing amplification \citep{Fuchs_01} and others are directly from simulations of galaxy discs \citep{Baba_13,Michikoshi_14}. We assume the pitch angles of our spirals to follow the following relation from \citet{Michikoshi_14}, taken from simulations. These predictions match up to analytical predictions from \citet{Fuchs_01} for $\Gamma < 1$, with the advantage that they cover the entire range of $\Gamma$ from 0--2. \citet{Michikoshi_14} note that the prediction is obtained from simulations of many-arm/flocculent structures, rather than grand design spirals. However, Fig.~2 of \citet{Michikoshi_14} shows that the model is well-matched to the observed spiral arm pitch angle of grand design spirals from \citet{Seigar_06} within $0.5 \lesssim \Gamma \lesssim 1.5$, where the majority of our spiral galaxies lie. We therefore apply this equation to all of the spiral galaxies in our sample. The predicted pitch angle is given by:\begin{equation}
\label{eq:michikoshi_psi}
\psi = \frac{2}{7} \frac{\sqrt{4-2\Gamma}}{\Gamma} . 
\end{equation} The value $\Gamma$ uses $\Omega$ and $\kappa$ defined in equations 4 of \citet{Donghia_15} and 6 of \citet{Donghia_13} for the Hernquist profile and \ref{eq:omega2_burkert} and \ref{eq:kappa2} of this paper for the Burkert profile. An issue that we have is deciding where to measure the spiral arm number; Fig.~2 of \citet{Bosma_99} clearly demonstrates that lower order modes are more strongly amplified in the inner regions, and higher order modes are more strongly amplified when one reaches the edge of the disc, an effect we see in Fig.~\ref{fig:typical_galaxies}. At small radii, the bulge, and potentially the presence of weak bars, makes arms hard to distinguish. At too large a distance from the galaxy centre, arms are too faint to distinguish. For the purpose of this paper, we choose to predict spiral arm number at 2 disc scale radii, a radius which should be well into the galaxy disc, yet far enough out that the inner features of a galaxy do not affect the measurement. The effect of measuring arms at different radii is discussed further in Sec.~\ref{sec:model_realism}. We also choose to predict spiral arm pitch angles at 2 scale radii in the rest of this paper.

\subsubsection{Predicted arm numbers for typical spirals}
\label{sec:typical_galaxies}

\begin{figure*}
\centering
\includegraphics[width=0.975\textwidth]{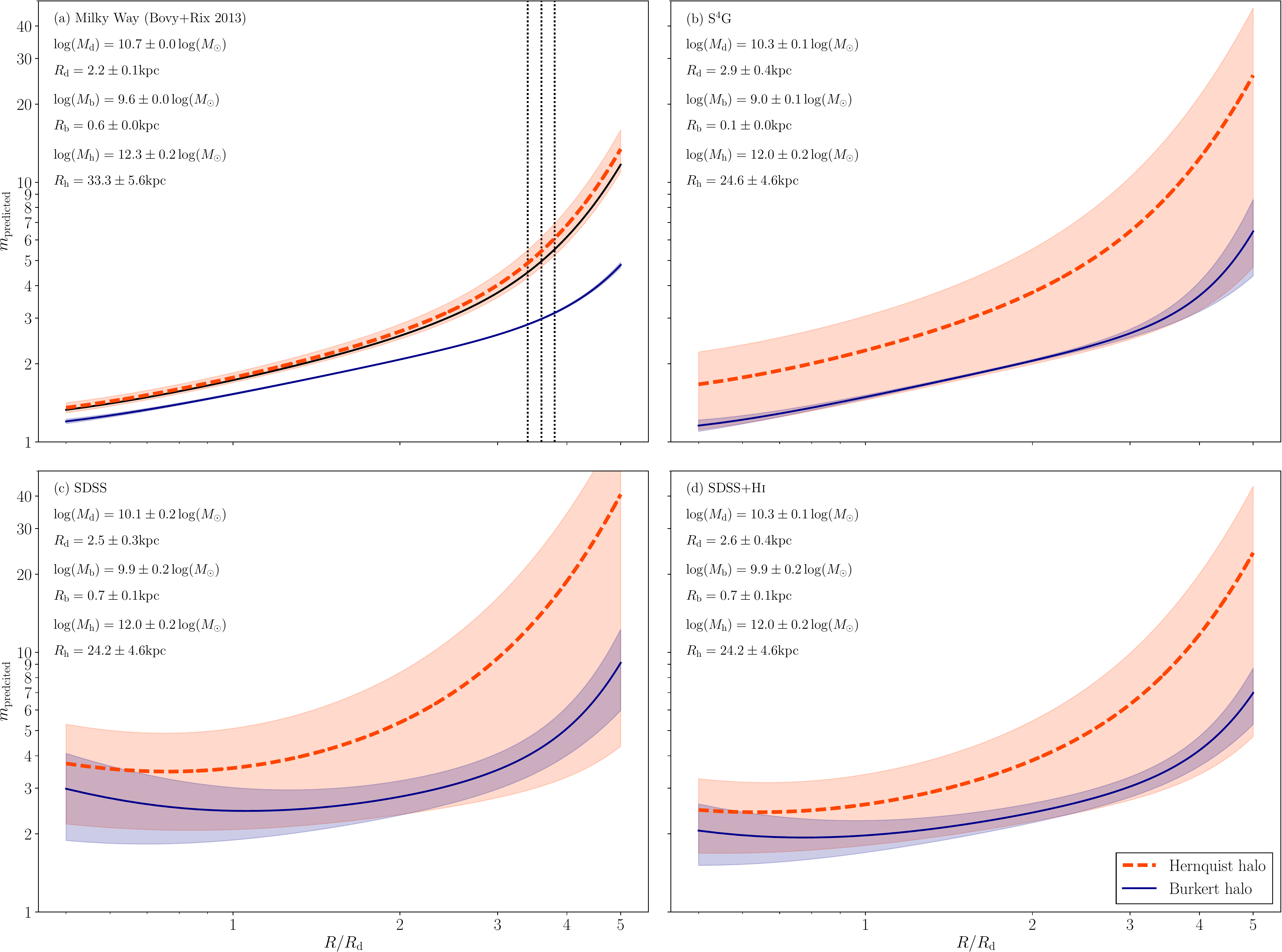}
\caption{Spiral arm number vs. radius for four typical spiral galaxies: (a) the Milky Way \citep{Bovy_13}; (b) a galaxy with median properties from the \sfg{} sample; (c) a galaxy with median properties from the SDSS sample and (d) a galaxy with median properties from the \sdsshi{} sample. The orange dashed lines show the expected spiral arm numbers for the cusped Hernquist dark matter profile, and the thinner, solid blue lines show the expected arm number for the cored Burkert profile. Both galaxy properties and halo shape have a strong influence on the expected spiral arm number in galaxies. The solid black line in (a) indicates the prediction for the Milky Way from \citet{Donghia_15}. The vertical black dotted lines in (a) indicate the radius of the solar neighbourhood, at $7.94 \pm 0.42$ kpc, or $3.6 \pm 0.2 R_\mathrm{d}$ \citep{Horrobin_04}. The disc, bulge and halo parameters are listed in the top-left corner of each sub plot.}
\label{fig:typical_galaxies}
\end{figure*}

The differences in halo profiles can have a strong influence on the expected spiral arm numbers in galaxies. In Fig.~\ref{fig:typical_galaxies}, the spiral arm numbers predicted from the galaxy model described in Sec.~\ref{sec:model} are shown for typical spiral galaxies from the \sfg{} and SDSS samples used in this paper. For reference, we also compute the halo properties of the Milky Way using the structural parameters of \citet{Bovy_13}. Their measured value of $5.9\pm0.5 \times 10^{10} \mathrm{M_\odot}$ predicts a halo of mass $\mathrm{M_{200}}=2.14 \pm 0.83 \times 10^{12} \mathrm{M_\odot}$ and scale radius $a_\mathrm{h}=34.5 \pm 4.5$ kpc. The predicted number of arms for the Milky Way for this halo mass, disc mass and a galaxy bulge of mass $4 \times 10^{9}\mathrm{M_\odot}$ and scale radius 0.6 kpc (as used in \citealt{Donghia_15}) are shown in Fig.~\ref{fig:typical_galaxies}a. We see a small offset in that the model predicts more spiral arms than the \citet{Donghia_15} Milky Way model -- this difference is due to the differences in the halo mass and size, with our predicted halo being larger in mass and than the one used in \citet{Donghia_15}.  

In Fig.~\ref{fig:typical_galaxies}b-d, we plot the same radius vs. predicted arm number trend for galaxies typical of the \sfg, SDSS and \sdsshi{} samples. We use the median values for disc, bulge and halo masses and sizes, and median error values for each sample. The variations in the galaxy parameters discussed in Sec.~\ref{sec:galaxy_properties} lead to changes in the expected spiral arm morphology. The median \sfg{} galaxy follows the trend of the Milky Way fairly closely, albeit with a larger error in the expected spiral arm number, owing to greater uncertainty in the measured bulge and disc parameters. A galaxy typical of the SDSS sample predicts more spiral arms, owing to the fact that the bulge is more prominent for this model -- this leads to an increase in the size of the bulge term in Eq.~\ref{eq:donghia_m}, which in turn increases the predicted spiral arm number. Including the \hi{} component in the SDSS model makes the disc more dominant, which leads to a suppression of the expected spiral arm number, which can be seen comparing Fig.~\ref{fig:typical_galaxies}c and d. We also see the direct influence that the dark matter profile shape has on the spiral arm numbers predicted for our galaxy model. The Hernquist profile is strongly cusped in the centre, whereas the Burkert profile is almost flat. The Burkert halo therefore has less influence on the spiral arm number in the baryon-dominated centre of galaxies, leading to systems being more disc dominated and therefore having fewer spiral arms. The predicted spiral arm number is also distinctly flatter in the inner regions, which is particularly apparent for the SDSS and \sdsshi{} samples in Fig.~\ref{fig:typical_galaxies}c-d. From these plots we can conclude that there are a number of factors that influence the spiral arm number in the model: more disc-dominated systems should have fewer spiral arms, and systems with flatter dark matter halo profiles should also have systematically fewer spiral arms.

The models outlined in this section give directly predictable arm numbers and pitch angles. All of the predictions are taken from direct analytical calculations of swing amplification theory and disc stability, and further verified by simulations. This simple galaxy model, with arm morphology predictions from only a bulge, disc and dark matter halo can now be tested with respect to observed visual galaxy morphology.

\section{Comparing model predictions with observations}
\label{sec:results}

In this section, we compare the predictions of swing amplification outlined in Sec.~\ref{sec:model} with observed morphologies of spiral galaxies. We begin by looking at the predicted arm number and pitch angle distributions from the Burkert and Hernquist haloes, in order to check whether they match the overall distributions we observe in real galaxies. We then look at how well the model can predict spiral arm numbers on a galaxy by galaxy basis, looking in more detail at the properties the dark matter halo requires for the model to work. 

\subsection{Spiral arm number distributions}
\label{sec:arm_number_distributions}

\begin{figure}
\centering
\includegraphics[width=0.45\textwidth]{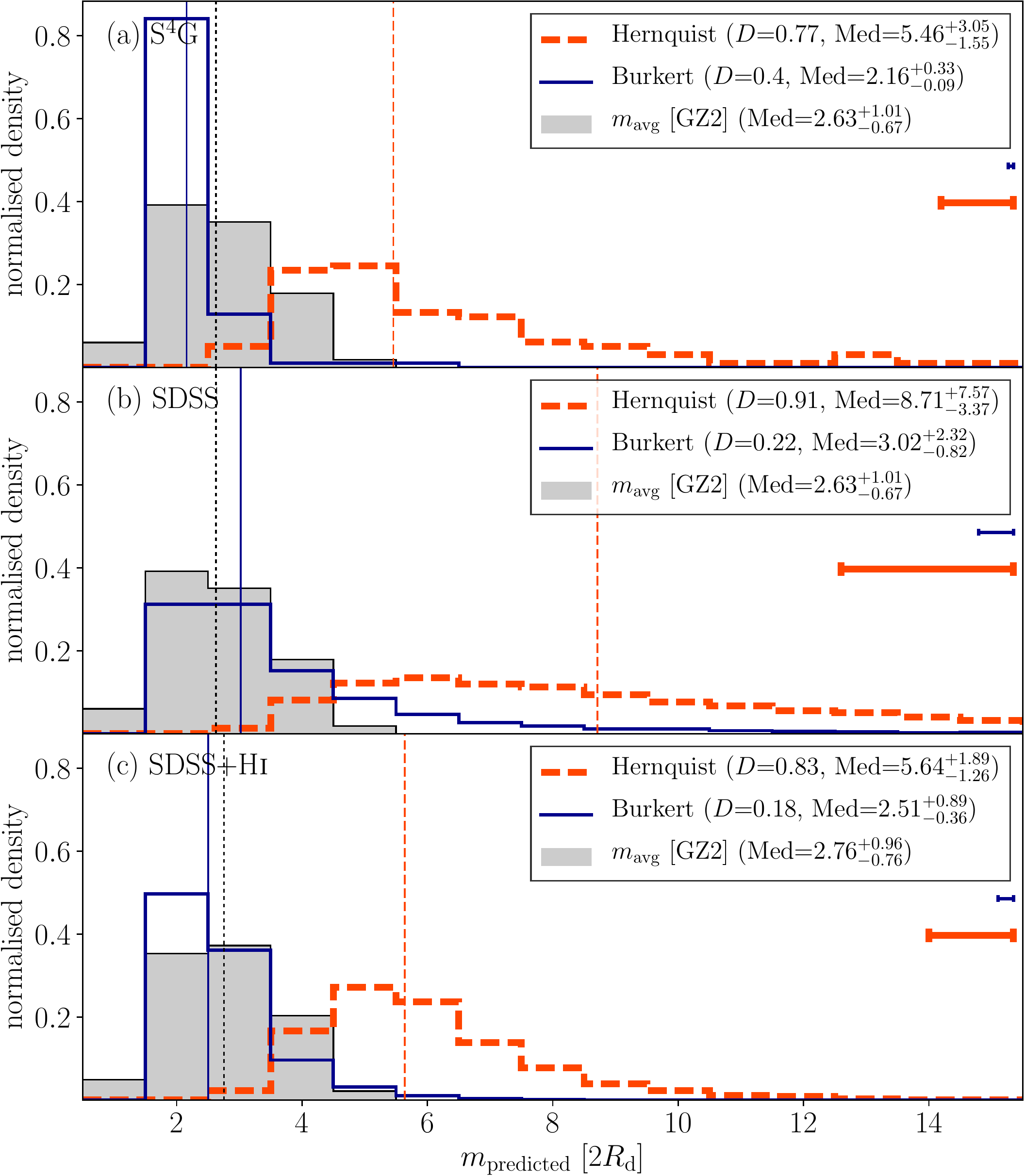}
\caption{Distributions of predicted spiral arm number for (a) \sfg{}, (b) SDSS and (c) \sdsshi{} galaxy samples. The grey histograms show the distributions of average arm number $m_\mathrm{avg}$ for the SDSS galaxy sample in a-b and the \sdsshi{} sample in c, and the vertical dotted black line shows their corresponding median values. The orange dashed histograms show the expected distribution for the Hernquist halo model, and the blue line shows the same distribution for the Burkert halo model. The error bars show the median error on the predicted $m$-value for each sample.}
\label{fig:m_distributions}
\end{figure}

Spiral arm numbers pose an interesting challenge to both observers and modellers of disc galaxies. From observations, we know that low arm numbers are preferred, with two-arm structures being particularly prevalent in the low-redshift Universe \citep{Elmegreen_82,Grosbol_04,Hart_16}. However, simulations often try to predict spiral arm numbers in the absence of bars. In this case, simulated spiral patterns are typically dominated by higher-order modes i.e. many-arm patterns (see \citealt{Dobbs_14} and references therein). 

We plot the distributions of spiral arm numbers for our samples of spiral galaxies in Fig.~\ref{fig:m_distributions}. The observed GZ2 $m_\mathrm{avg}$ arm number distribution of the SDSS sample is plotted for reference in each panel. Additional histograms show the arm number distributions predicted by our model for each halo type and sample. Fig.~\ref{fig:m_distributions}a shows the \sfg{} sample, Fig.~\ref{fig:m_distributions}b shows the SDSS sample and Fig.~\ref{fig:m_distributions}c shows the \sdsshi{} sample. In Fig.~\ref{fig:m_distributions}a, the SDSS sample is used for comparison, given its similarity in total stellar mass. From the observed spiral arm numbers, we see the familiar trend that disc galaxies tend to prefer lower order spiral modes, with the two-arm mode being particularly prevalent -- the modal bin is centred on $m_\mathrm{avg}=2$, and the median arm number is \upperlower{2.63}{1.01}{0.67}, where the $\pm$ values denote the 16th and 84th percentiles. For \sdsshi{}, the modal bin is centred on 2.5 and the distribution has median arm number \upperlower{2.76}{0.96}{0.76}. The galaxy model with the Hernquist halo clearly produces too many spiral arms, with median arm number \upperlower{5.46}{23.05}{1.55} for the \sfg{} sample, \upperlower{8.71}{7.57}{3.37} for the SDSS sample and \upperlower{5.65}{1.89}{1.27} for the \sdsshi{} sample. The reasons for these differences between the samples were discussed in Sec.~\ref{sec:typical_galaxies}. We can quantify how closely related these distributions are using the KS $D$-statistic.\footnote{Our simplified model is unlikely to recover the range of morphologies exactly, so the KS $p$-value is likely to converge to close to 0 in all cases, making it unsuitable for distinguishing any differences.} If one instead models the distributions with a cored Burkert dark matter halo (thinner blue lines), we see that the distributions of spiral arm number match a realistic spiral arm number distribution more closely, with median $m_\mathrm{avg}$-values of \upperlower{3.02}{2.32}{0.82} for the pure stellar sample and \upperlower{2.51}{0.89}{0.36} for \sdsshi{} and much lower $D$-statistics of 0.22 and 0.18 respectively. The result for the \sfg{} sample in Fig.~\ref{fig:m_distributions}a is that we produce too many two-arm galaxies. We note, however, that the comparison is less certain, given the different sample selections for \sfg{} and SDSS, and the potential discrepancies discussed in Sec.~\ref{sec:galaxy_properties}.

\subsection{Spiral arm pitch angle distributions}
\label{sec:pitch_angle_distributions}

\begin{figure}
\centering
\includegraphics[width=0.45\textwidth]{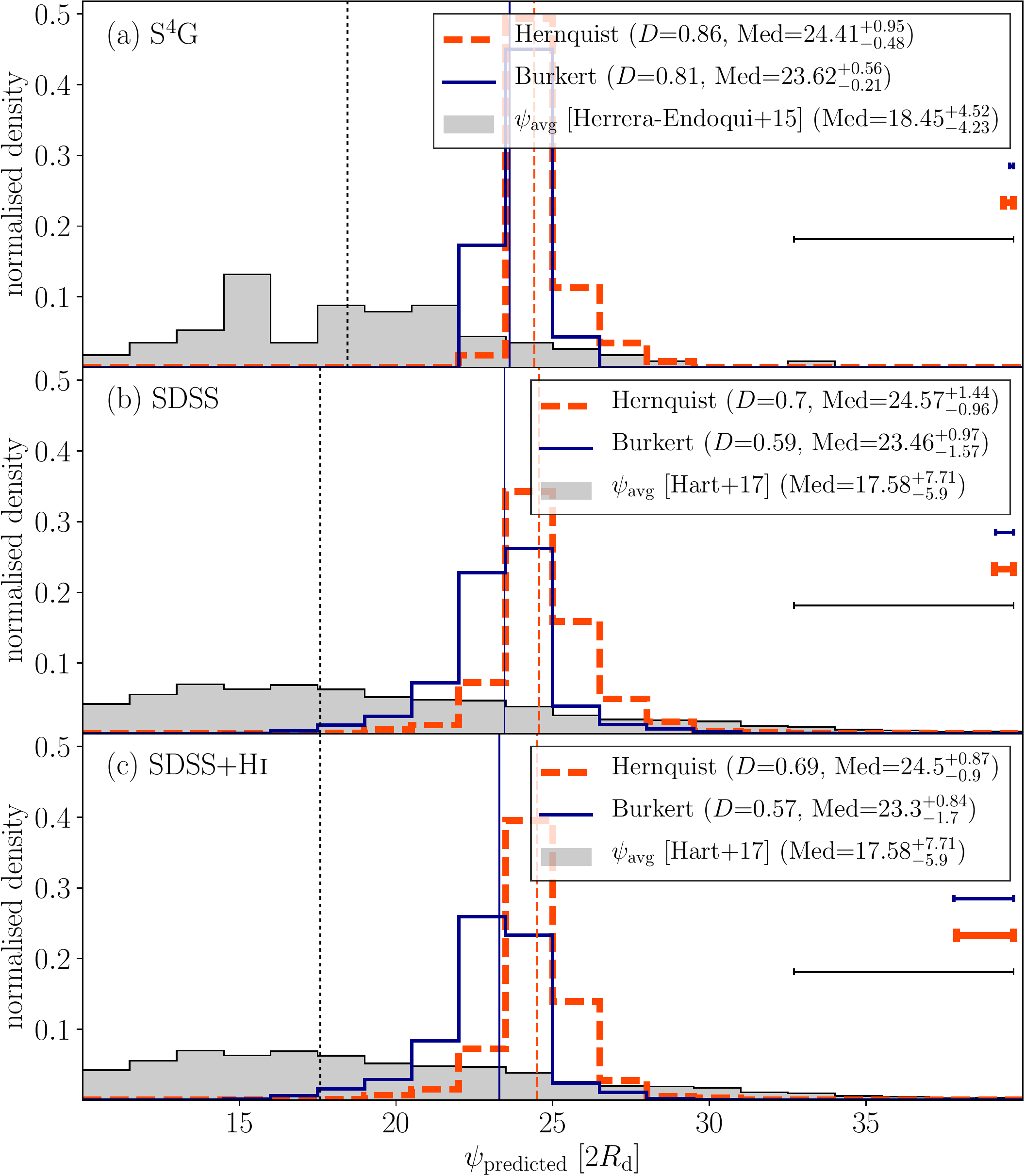}
\caption{Distributions of predicted pitch angle for (a) \sfg{}, (b) SDSS and (c) \sdsshi{} galaxy samples. The grey filled histograms show the measured spiral arm pitch angle distributions for each sample from \citet{Herrera-Endoqui_15} for \sfg{} and \citet{Hart_17b} for the SDSS and \sdsshi{} samples. The orange dashed histograms show the expected distribution for swing amplified arms, assuming the Hernquist halo model, and the blue line shows the same distribution for the Burkert halo model. The median error in each measurement is shown by the error bar in each panel and the black error bar shows the estimated observational error from \citet{Hart_17b}.}
\label{fig:psi_distributions}
\end{figure}

The spiral arm pitch angle measures how tightly wound spiral arms are. The expected pitch angle in spiral galaxies depends on the underlying mass distribution, with more centrally concentrated masses leading to tighter spiral arms. This is usually predicted to be the case, no matter which mechanism is responsible for producing the arms \citep{Fuchs_01,Seigar_08}. However, other properties such as the age of the spiral arm \citep{Grand_12} and the number of arms \citep{Hart_17b} can affect pitch angles. From the simulations of \citet{Michikoshi_14}, we can directly predict the pitch angle given the rate of shear in the disc of a galaxy (see Sec.~\ref{sec:model}). We plot the expected distributions of spiral arm pitch angle in Fig.~\ref{fig:psi_distributions}. The grey distributions show the observed spiral arm pitch angles measured for the \sfg{} sample and the SDSS samples from \citet{Herrera-Endoqui_15} and \citet{Hart_17b} respectively. If the model perfectly fit the spiral galaxy population as a whole, one would expect a distribution of pitch angles centred on $\sim$19\textdegree{} and 16th-84th percentile range of $\sim$12-15\textdegree{} for each sample. Instead, for each dark matter halo profile, we observe a narrow range of pitch angles with looser spiral arms (larger pitch angles).

\begin{figure}
\centering
\includegraphics[width=0.45\textwidth]{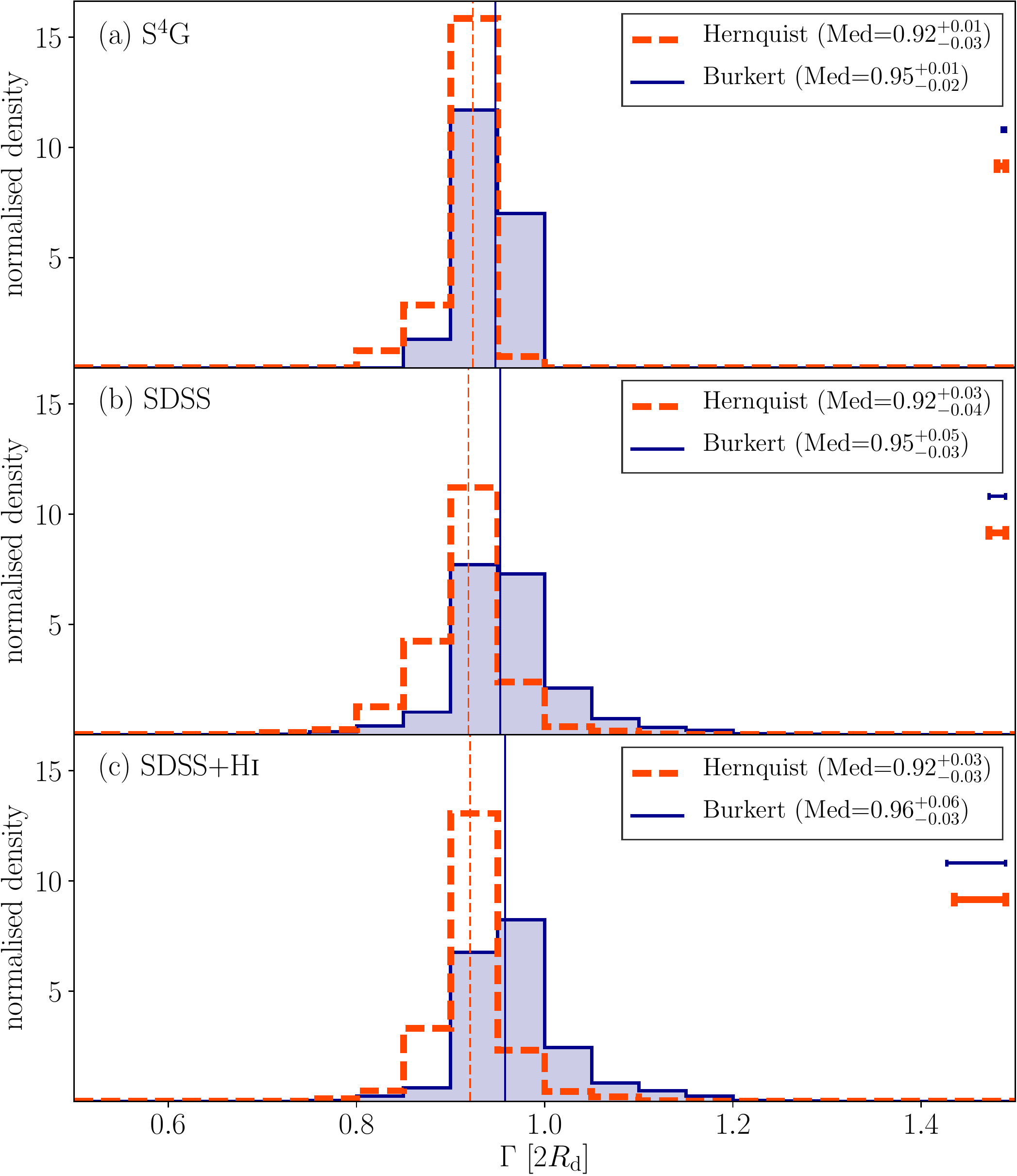}
\caption{Distributions of predicted values for shear, $\Gamma$, for (a) \sfg{}, (b) SDSS and (c) \sdsshi{} galaxy samples. The filled blue histograms show the values with a Burkert halo, and the dashed orange histograms show the distributions with a Hernquist halo. The median error in each measurement is shown by the error bar in each panel.}
\label{fig:Gamma_distributions}
\end{figure}

The Burkert profile leads to spiral arms which are tighter than those in the Hernquist profile, but leads to distributions which are peaked at $\sim24$\textdegree{}. Fig.~\ref{fig:Gamma_distributions} shows the distributions of the shear, $\Gamma$. Both the Hernquist and the Burkert halo in our galaxy model predict $\Gamma \sim$1. The Hernquist profile has distributions of lower $\Gamma$ values, but neither model gives distributions of $\Gamma$>1 required to produce the distributions of tighter spiral arms observed in real spiral galaxies. 

One potential reason for the discrepancies in the pitch angles is measurement error. In \citet{Hart_17b}, we derived two alternative pitch angle measurements, and saw a scatter of $\approx 7$ \textdegree{}. Convolving the predicted pitch angle distributions with a random Gaussian error of 7\textdegree{} leads to the widening of the distributions -- the 16th-84th percentile range is $\approx12$\textdegree{} for the \sfg{} sample and $\approx15$\textdegree{} for the SDSS samples in this case. This can account for the discrepancy between the measured and observed pitch angles. However, the peaks of the predicted pitch angle distributions are still too loose compared to those observed. 

\begin{figure}
\centering
\includegraphics[width=0.45\textwidth]{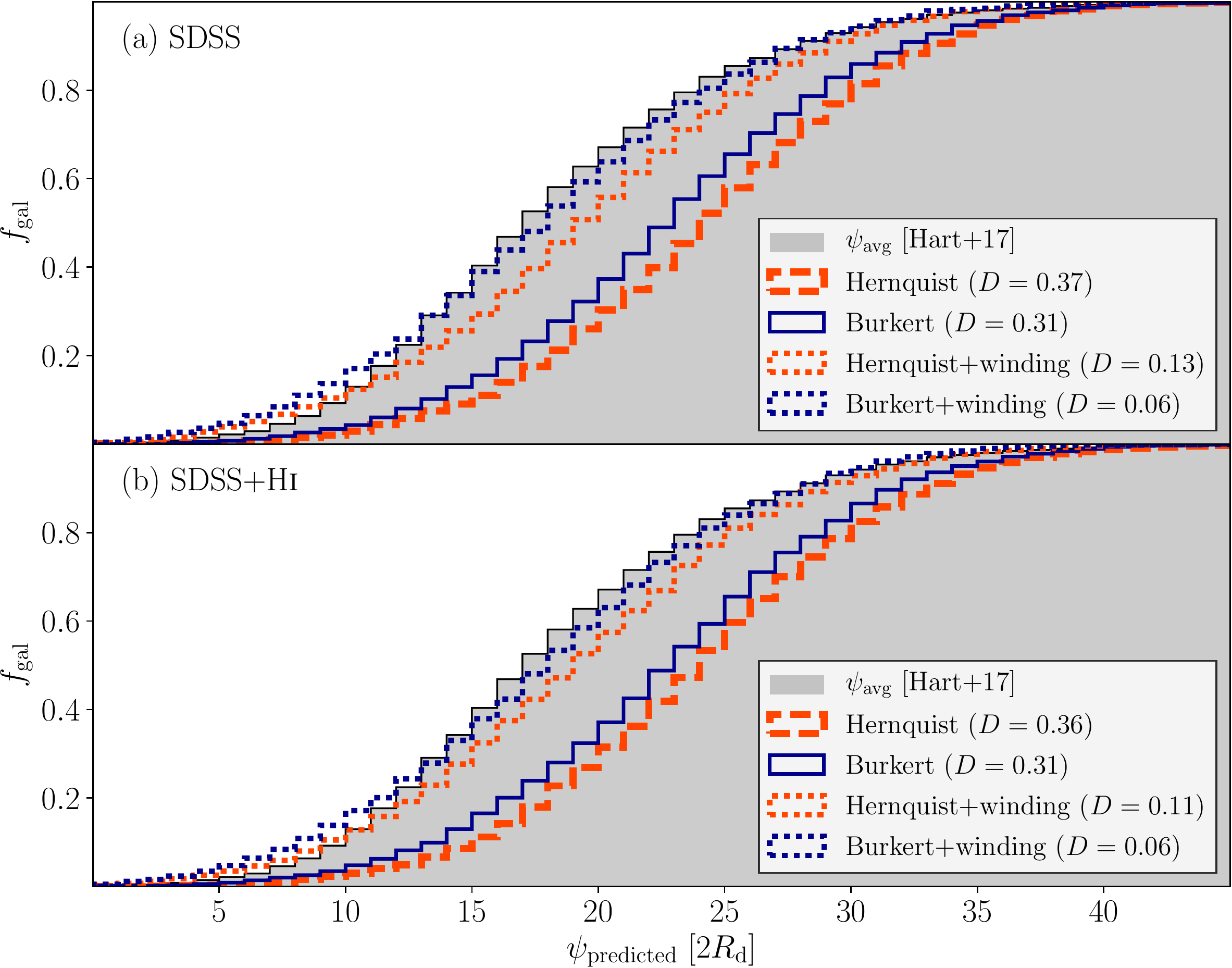}
\caption{Cumulative fractions of observed (grey filled histograms) and predicted spiral arm pitch angles (stepped histograms) for (a) the SDSS sample and (b) the SDSS+\hi{} sample. The dashed orange line and the solid blue line show the distributions of pitch angles from the model using Hernquist and Burkert dark matter profiles respectively, convolved with a Gaussian error of 7\textdegree{}. The dotted orange and blue lines indicate the same distributions, with a random scatter downwards of 0-10\textdegree{}. This scatter makes the predicted pitch angles match the real distribution more closely.}
\label{fig:cumulative_psis}
\end{figure}

In Fig.~\ref{fig:cumulative_psis}, we show the cumulative distributions of spiral arm pitch angles for the model compared with the observations, with the predictions convolved with 7\textdegree{} errors. The picture which emerges is interesting -- the maximum pitch angle seems to be the same between the observations and predictions. The 99.7th percentile is $\psi=44.3$\textdegree{} in the observations; the equivalent values are 43.3, 44.8, 45.0 and 43.0\textdegree{} for the SDSS with the Hernquist halo, SDSS with the Burkert halo, SDSS+\hi{} with the Hernquist halo and the SDSS+\hi{} with the Burkert halo respectively. However, the model deviates from the observed distributions for tighter spiral arms. Swing amplified arms are, however, material in nature, and do wind up over time. \citet{Grand_13} demonstrated that spiral arms exist for $\approx100$ Myr, and wind up by approximately 10\textdegree{} over the course of their lifetime. The dotted lines in Fig.~\ref{fig:cumulative_psis} show the same galaxies, with a random winding of 0-10\textdegree{} applied to each galaxy (each galaxy is randomly 0-10\textdegree{} tighter than the model prediction). In this case, we see the model is much more consistent with the observations. This is particularly the case for the Burkert dark matter profile, where the KS $D$-statistic has been reduced to $\leq 0.1$ in both the SDSS and SDSS+\hi{} cases. In order to match the distributions correctly, the winding up of spiral arms must also be taken into account.

These results show that spiral arm number is the better diagnostic tool for finding swing amplified spiral modes. The model inputs that we employ cannot reproduce the subtle differences in $\Gamma$ that are required to produce the variety of pitch angles in galaxies. The errors on the measured pitch angles are also relatively large, of order 7\textdegree{} \citep{Hart_17b}. This makes any model difficult to constrain due to a large scatter introduced by the errors in the measurements. Additionally, the age of the spiral arms appears to have an effect -- if spiral arms do wind up over time, as the evidence here suggests, then this will introduce an unwanted, difficult to quantify scatter in the true pitch angles of spiral galaxies.

\subsection{The disc fraction-arm number relation}
\label{sec:fd_vs_m}

\begin{figure*}
\centering
\includegraphics[width=0.975\textwidth]{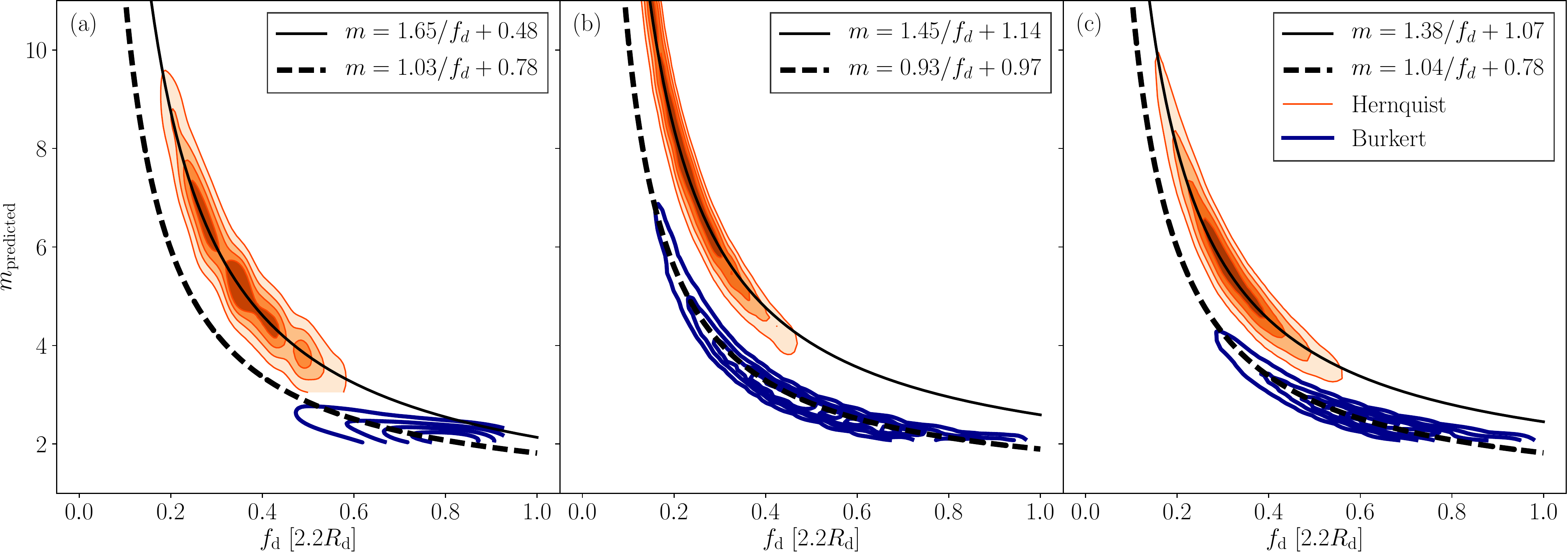}
\caption{Disc fraction $f_\mathrm{d}$ vs. predicted spiral arm number $m_\mathrm{predicted}$ for (a) the \sfg{} sample, (b) the SDSS sample and (c) the \sdsshi{} sample. The orange filled contours show the predictions for the Hernquist halo and the blue lined contours show the predictions for the Burkert halo. The contour lines show where 20, 40, 60 and 80 per cent of the data lie for each sample. The flatter inner profile leads to fewer predicted arms for a given disc fraction.}
\label{fig:fd_vs_m}
\end{figure*}

From the formalism described in Sec.~\ref{sec:model}, the predicted spiral arm number is expected to have a strong dependence on the relative sizes and masses of haloes, bulges and discs. Of particular note is the relation with the disc fraction: many simulations predict a strong correlation of $m \propto f_\mathrm{d}^{-1}$, where $f_\mathrm{d}$ is the disc fraction within 2.2 times the disc scale length \citep{Carlberg_85,Bottema_03,Fujii_11,Donghia_13}. Such a relation is unsurprising, given the functional form of Eq.~\ref{eq:donghia_m}. The equation has terms predicting $m \propto \mathrm{M_b/M_d}$ and $m \propto \mathrm{M_h/M_d}$; the only complications are the other dependencies on the relative sizes of the components. In Fig.~\ref{fig:fd_vs_m}, the relation for each of the sub samples is shown. Here we see the expected relationship of $m_\mathrm{predicted} \propto f_\mathrm{d}^{-1}$. The scatter is very small, meaning the relationship is dominated by the mass fractions, rather than the  relative sizes of the components. We also see another trend that the relationship depends not only on $f_\mathrm{d}$, but the shape of the dark matter profile also plays a role: the Burkert profile, which has a flat inner region, leads to a lower predicted spiral arm number for a given disc fraction as well as larger disc fractions. 

\subsection{Predicting spiral arm numbers in galaxies}
\label{sec:predicting_arm_numbers}

\begin{figure}
\centering
\includegraphics[width=0.45\textwidth]{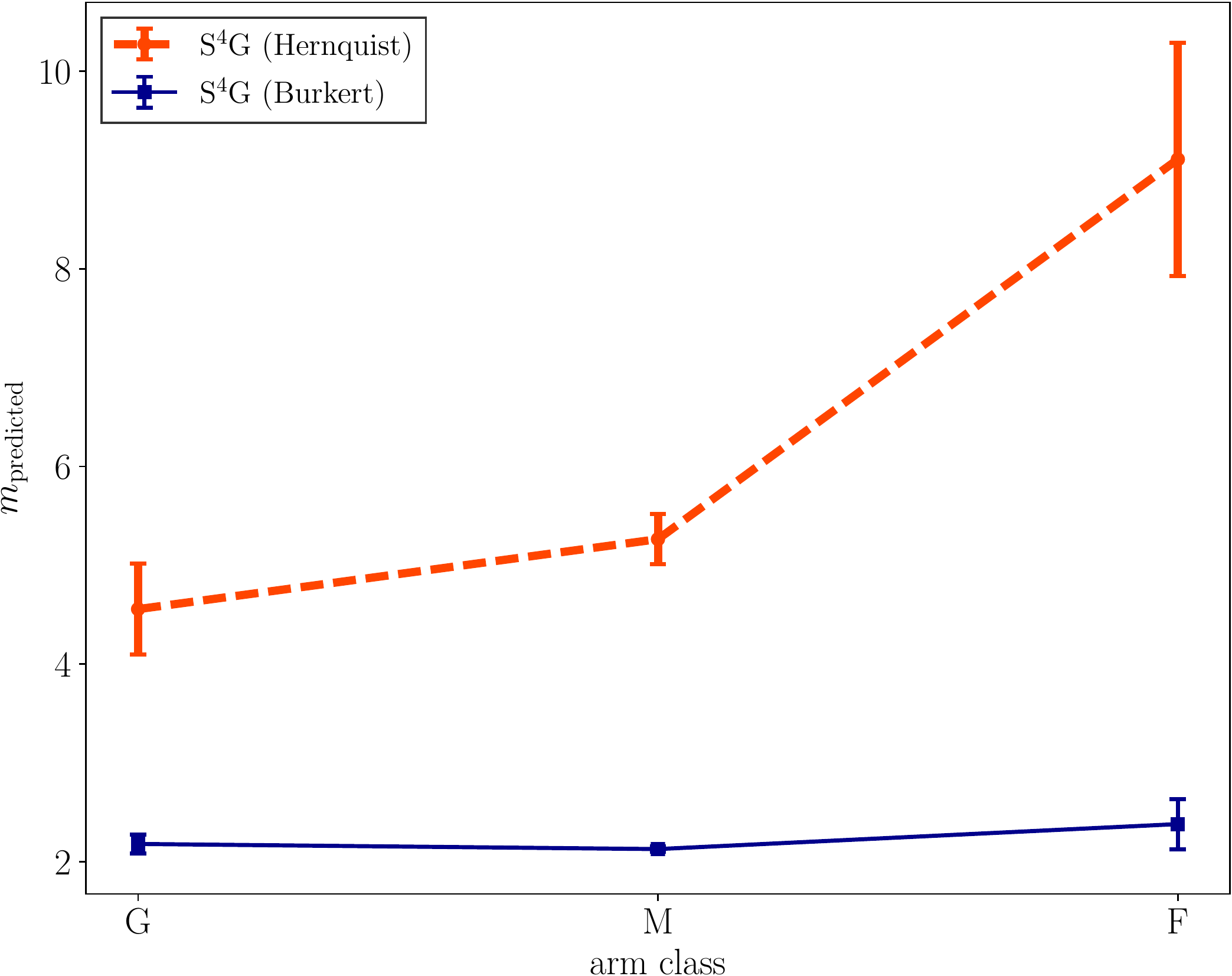}
\caption{Predicted spiral arm numbers for \sfg{} grand design (G), many-arm (M) and flocculent (F) spirals. The orange dashed line shows the median spiral arm number for the Hernquist dark matter halo, and the solid blue line shows the same value for the Burkert dark matter halo. The error bars indicate one standard error on the median for each sub sample.}
\label{fig:m_vs_m_s4g}
\end{figure}

\begin{figure}
\centering
\includegraphics[width=0.45\textwidth]{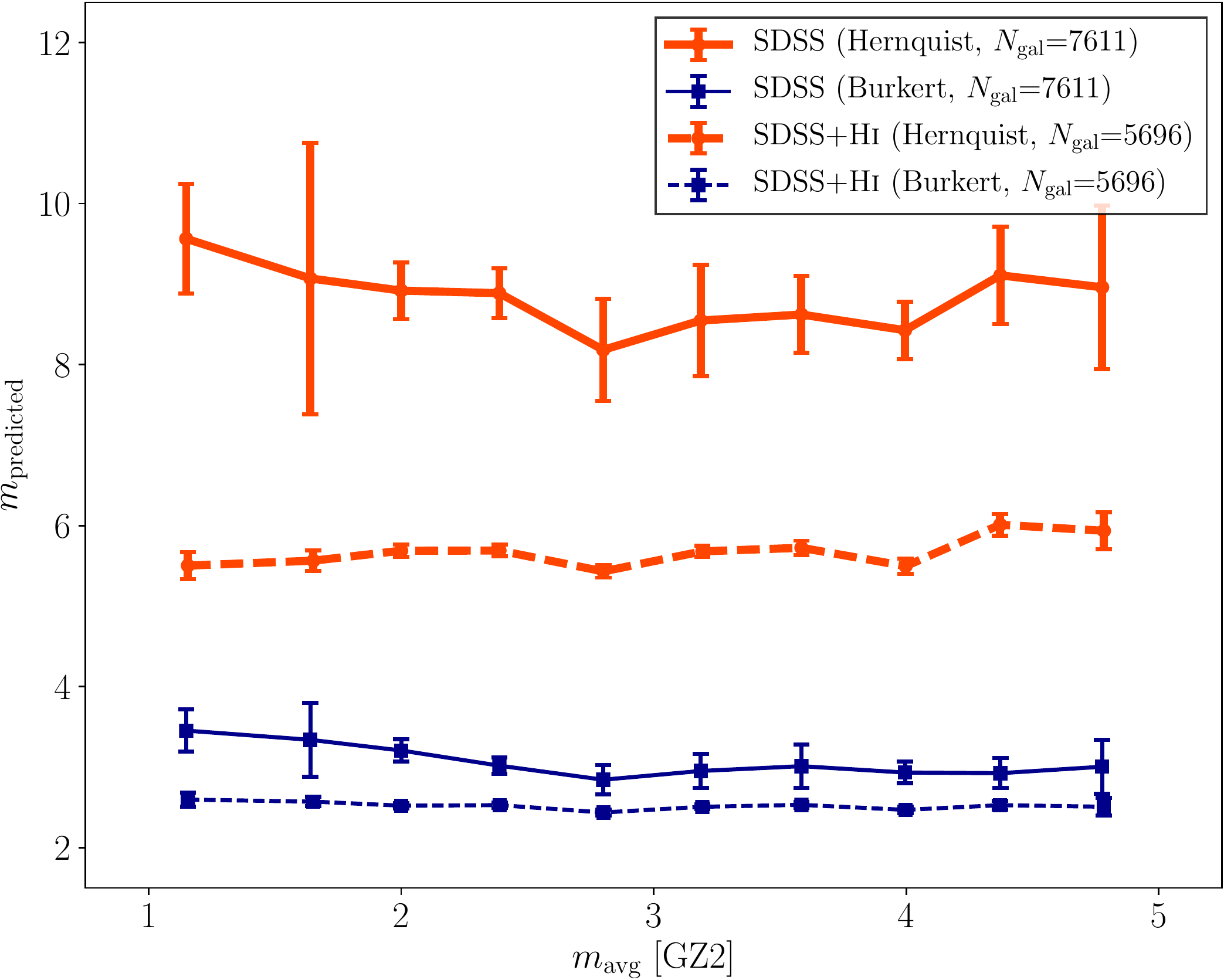}
\caption{Spiral arm number measured from GZ2 $m_\mathrm{avg}$ vs. spiral arm number predicted from the galaxy model. The orange lines show the models with a Hernquist dark matter halo, and the blue lines show the models with a Burkert dark matter halo profile. The solid lines show the SDSS samples, and the dashed lines show the \sdsshi{} samples. The points show the median and the error bars indicate one standard error on the median. There is no clear correlation to confirm that the model can predict spiral arm numbers accurately with either a cusped or cored halo.}
\label{fig:m_vs_m}
\end{figure}

Given that the modal spiral arm theory does seem able to predict reasonable spiral arm number distributions, given a cored dark matter profile, we will now investigate how well the theory predicts spiral arm numbers in individual galaxies. If the modal theory is indeed accurate, we expect to see a strong correlation between the observed spiral arm numbers and those predicted by Eq.~\ref{eq:donghia_m}. 

As a first test, we check the predicted spiral arm numbers for the \sfg{} sample. This sample is observed in the near infra-red, so the spiral arms we see here should correspond to the underlying mass distributions of the spiral galaxies. For validation of our SDSS results, we use the already published arm classifications for the \sfg{} galaxies from \citet{Buta_15}. Galaxies are classified by their Elmegreen arm-type, as either grand design, many-arm or flocculent. Grand design (G) spiral galaxies are characterised by their strong two-arm structure, whereas many-arm (M) spirals instead have more than two spiral arms, and flocculent (F) spiral galaxies have more, broken, patchy spiral arms than many-arm galaxies \citep{Elmegreen_82}. From these arguments, we expect the grand design spiral galaxies to have the fewest predicted spiral arms, and the flocculent galaxies to have the most predicted spiral arms. In Fig.~\ref{fig:m_vs_m_s4g}, the median predicted spiral arm number is shown for each of the spiral arm subcategories. There is a weak trend for exactly what we expect: the flocculent spirals do have the most predicted spiral arms, with $m_\mathrm{predicted}=9.1\pm1.2$. There is, however, little difference between the grand design and many-arm spiral categories with $m_\mathrm{predicted}=4.6\pm0.5$ and $5.3 \pm 0.3$. We also see evidence that a cored Burkert dark matter halo profile cannot reproduce the variability in spiral structure between spiral galaxies -- in all cases, the predicted spiral arm number is $\sim 2$.

For the SDSS sample, we have direct measurements of spiral arm numbers from the GZ2 classifications of spiral galaxies. Rather than asking questions to describe the spiral arm type, Galaxy Zoo volunteers instead classified the number of spiral arms they could observe in the optical image. We use the average arm number $m_\mathrm{avg}$ to describe the spiral arm number for each galaxy. The number of spiral arms observed for the SDSS and \sdsshi{} vs. the number of spiral arms predicted for those same galaxies are shown in Fig.~\ref{fig:m_vs_m}. Here, there is no evidence that the optically classified spiral arm number has any relation to the number of spiral arms predicted. This is the case for the cusped Hernquist and cored Burkert profiles, with and without the disc gas mass being included in the prediction. We observe no strong correlation between the predicted and observed spiral arm numbers, with $\vert r_s \vert \leq 0.1$ in each case.

\subsection{Varying the dark matter halo}
\label{sec:varying_halo}

Neither a cored or cusped halo can produce the variety of spiral arm morphologies in local galaxies from grand design to many-arm structures for our samples. This does not necessarily mean spiral arms are not swing amplified modes; instead, the dark matter halo may exhibit strong differences from galaxy to galaxy. In fact, the radial profiles of dark matter haloes have been shown to vary greatly from galaxy to galaxy, with earlier type massive ellipticals having cuspier profiles \citep{Dutton_13,Dutton_14,Sonnenfeld_15}, and later type, low surface brightness systems having flatter inner dark matter profiles \citep{deBlok_01,Swaters_03,Goerdt_06}, with some level of interpolation in between \citep{Dutton_16}. 

With the mathematics formulated in Sec.~\ref{sec:dark_matter_haloes} and Sec.~\ref{sec:swing_amplification_quantities}, we can interpolate between the Burkert (cored) and Hernquist (cusped) dark matter profiles. Models of dark matter haloes usually describe the profile shape in terms of halo contraction. Contracted haloes have less mass in their inner regions, and more in their outer regions, with star-formation feedback often cited as the cause of such a change \citep{Navarro_96,Oh_11,Katz_17}. In our model, we are principally concerned with the inner region of the halo. We can mimic the halo contraction in the inner region by varying the shape of the halo as described in Sec.~\ref{sec:halo_shapes}. 

In order to address the issue of whether an interpolated halo can reproduce the predictions of swing amplified spiral arms, we can ask the question of what value of $\alpha$ (a proxy for the contraction in the inner regions) our haloes need to be in order for a model to match perfectly. That Eq.~\ref{eq:donghia_m} can be split into multiple parts allows for easy manipulation when we consider our superimposed hybrid dark matter haloes described in Sec.~\ref{sec:halo_shapes}. The equations become:\begin{equation}
\label{eq:m_alpha}
\begin{aligned}
m(r) &= m_\mathrm{b}(r) + m_\mathrm{d}(r) + (1+\alpha) m_\mathrm{h,B}(r) &(\alpha < 0) \\
m(r) &= m_\mathrm{b}(r) + m_\mathrm{d}(r) + (1-\alpha)m_\mathrm{h,B}(r) + \alpha m_\mathrm{h,H}(r) &(0 \leq \alpha \leq 1) \\
m(r) &= m_\mathrm{b}(r) + m_\mathrm{d}(r) + \alpha m_\mathrm{h,H}(r) &(\alpha > 1) ,
\end{aligned}
\end{equation} where $m_\mathrm{h,H}$ and $m_\mathrm{h,B}$ are the Hernquist and Burkert halo arm numbers. We now have a set of inferred dark matter halo profile shapes, $\alpha$, based on each galaxy's measured bulge and disc properties, observed spiral arm number, and the assumption of the swing amplification model. Dark matter halo expansion is often quantified as the mass of the halo inside a given radius when baryonic processes have been taken into account divided by the mass the halo would have if only dark matter were present (e.g. \citealt{Dutton_16}). To this end, we define the following to estimate the same quantity:\begin{equation}
\label{eq:epsilon}
\epsilon_\mathrm{halo}  = \log(\mathrm{M_{halo}} [0.01R_{200}] / \mathrm{M_\mathrm{hernquist}} [0.01R_{200}]) ,
\end{equation} where $\mathrm{M_{halo}}$ is the mass of a given halo constructed from the superposition of the Hernquist and Burkert profile as described in Sec.~\ref{sec:halo_shapes}, and $\mathrm{M_\mathrm{hernquist}}$ is the mass of the Hernquist halo of the same mass and size. The Hernquist halo should approximate a dark matter only halo, given that dark matter simulations predict cuspy NFW-like haloes \citep*{Navarro_96,Navarro_97}. 

\begin{figure}
\centering
\includegraphics[width=0.45\textwidth]{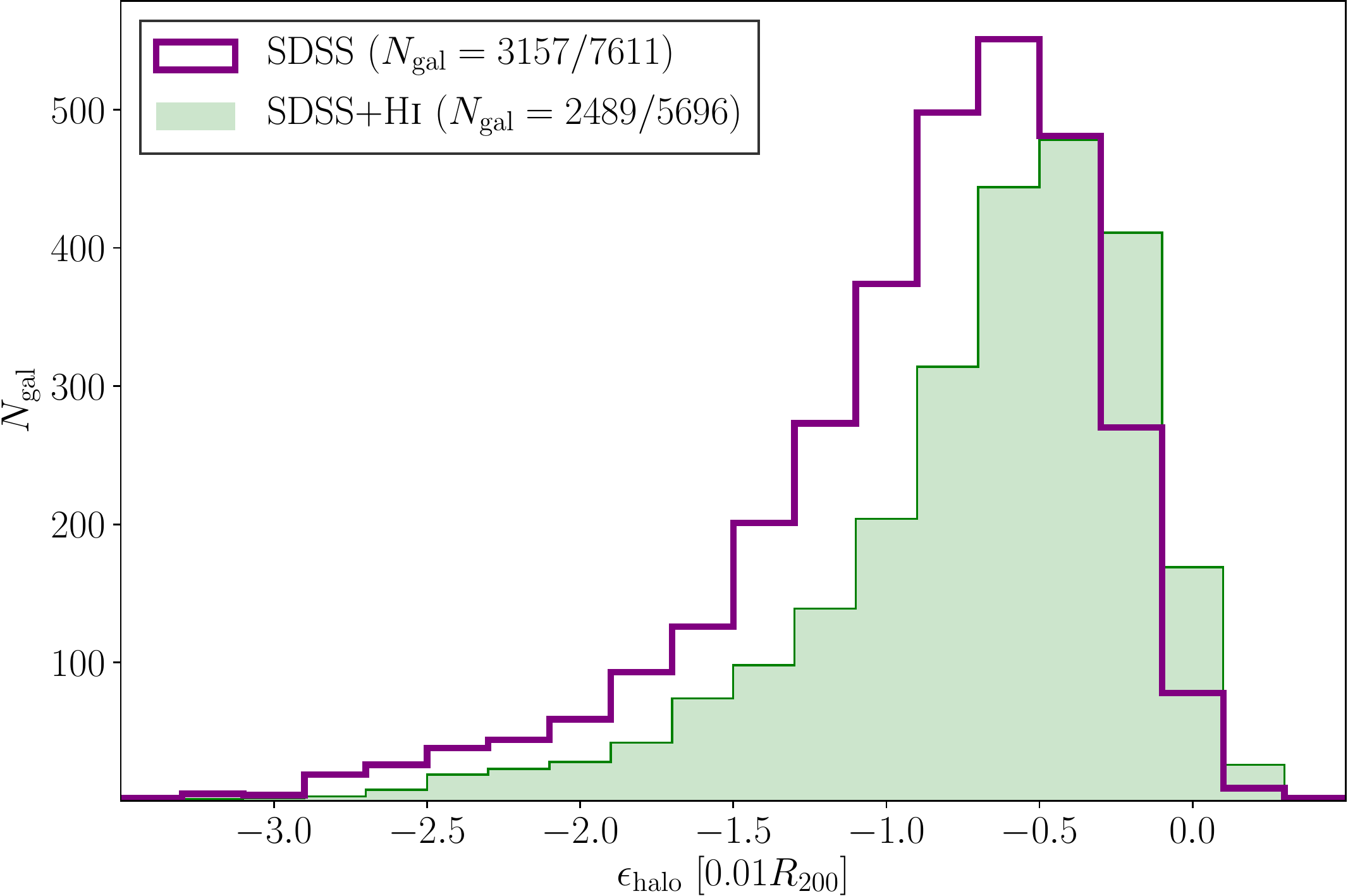}
\caption{Values of halo expansion parameter required to reproduce the spiral arm numbers from GZ2. The green filled histogram shows the \sdsshi{} sample and the purple stepped histogram shows the SDSS sample. A significant fraction of both populations cannot be explained by swing amplification, even with a dark matter halo set to $\mathrm{M_h} = 0$.}
\label{fig:epsilon_required}
\end{figure}

For each galaxy, we calculate the mass of the modified halo that gives the correct arm number vs. the mass of the halo one would expect if there were no baryonic processes affecting the halo. The distributions of the required $\epsilon_\mathrm{halo}$ values are shown in Fig.~\ref{fig:epsilon_required}. Only galaxies with physical dark matter haloes are included in this distribution -- these are galaxies with $\alpha>-1$, and make up 3157/7611 of the SDSS galaxies (41.5 per cent) and 2489/5696 of the SDSS+\hi{} galaxies (43.7 per cent). For the remaining $\approx 60$ per cent of the galaxies, the spiral arm number from the disc and bulge is greater then the observed spiral arm number; even in the extreme case where there is no dark matter halo contribution, the predictions cannot match the observations. Their origin is therefore unlikely to be swing amplification. These are usually galaxies with low spiral arm numbers: 71.3 per cent of these galaxies have spiral arm numbers of $m=1$ or $m=2$ according to GZ2. For the galaxies that may have swing amplified arms ($\alpha>-1$), this value is just 25.2 per cent.

\begin{figure*}
\centering
\includegraphics[width=0.975\textwidth]{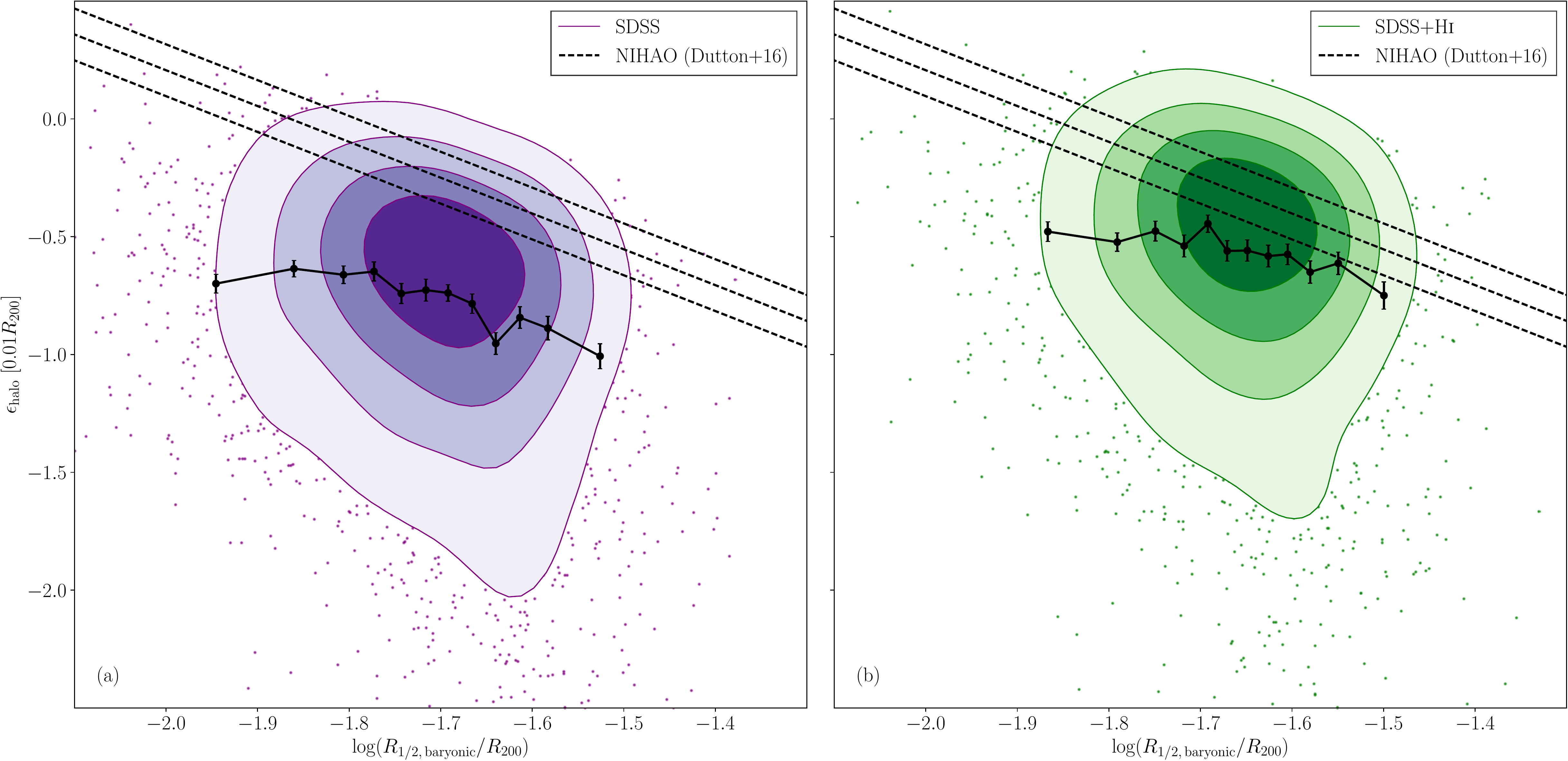}
\caption{Mass of the halo calculated from the swing amplification model divided by the mass of a cuspy Hernquist dark matter halo inside 0.01$R_{200}$. (a) shows the SDSS sample and (b) shows the \sdsshi{} sample. The contours enclose 20, 40, 60 and 80 per cent of the data points in each panel. Galaxies with physical ($\alpha \geq -1$) haloes lie in the region where one would expect to observe them if their spiral arms are swing amplified modes. The dashed lines show the prediction for the halo expansion/contraction from NIHAO \citep{Dutton_16}. The black points joined by a solid line show the median and the error bars indicate one standard error on the median.}
\label{fig:r_vs_epsilon}
\end{figure*}

To test whether these $\alpha$ parameters derived directly from observed quantities are reasonable, we compare them to results from simulations. The dark matter only halo mass, $\mathrm{M_{DMO}}$, is taken as simply the mass of the Hernquist halo. The mass of the halo required, $\mathrm{M_{h,req}}$, is the mass of the interpolated halo. Recent simulations have predicted that the size of the dark matter halo depends on a number of parameters related to the host galaxy \citep{DiCintio_14a,DiCintio_14b}. Notably, \citet{Dutton_16} simulated a range of galaxies with the NIHAO simulation suite, finding clear correlations between galaxy star formation efficiency, stellar mass and halo mass and the dark matter halo contraction. They also published a relationship between galaxy size, halo size and the halo contraction of the following form: \begin{equation}
\label{eq:nihao_contraction}
\begin{split}
\log(\mathrm{M^{0.01}_{hydro}/M^{0.01}_{DMO}}) &= -0.28(\pm0.11) \\ &- 1.52 (\pm 0.42)(\log(R_{1/2}/R_{200}) + 1.68) ,
\end{split}
\end{equation} where $\mathrm{M_{hydro}}$ is the mass of the dark matter halo simulated with baryonic processes and $R_{1/2}$ is the galaxy half mass radius. The superscripts 0.01 denote the mass within 0.01$R_{200}$, the central part of the halo where there is significant baryonic mass content: the term $\log(\mathrm{M^{0.01}_{hydro}/M^{0.01}_{DMO}})$ is therefore directly replaceable with our $\epsilon_\mathrm{halo}$ term. This correlation is used as a direct comparison for the data in this paper. The calculated dark matter halo contraction parameter vs. galaxy half light radius for both the SDSS and the \sdsshi{} samples are shown in Fig.~\ref{fig:r_vs_epsilon}. The dashed line defined by Eq.~\ref{eq:nihao_contraction} is also shown for reference. The majority of the galaxies lie to the right of $R_\mathrm{1/2} / R_{200}=2$, which is where the efficiently star-forming NIHAO disc galaxies lie, which is expected given that we consider star-forming spiral galaxies. Here, an interesting result emerges -- galaxies which have physical $\alpha$ values ($\alpha \geq -1$) require dark matter haloes very similar to the ones which the \citet{Dutton_16} simulations predict. The inclusion of a gas component also appears to bring the overall distributions closer to where one would expect the spiral galaxy sample in this paper to lie. There also appears to be a negative correlation between halo expansion and baryonic-to-halo size in both cases, as indicated by the solid black line in each panel, in agreement with the NIHAO simulation.

\begin{figure}
\centering
\includegraphics[width=0.45\textwidth]{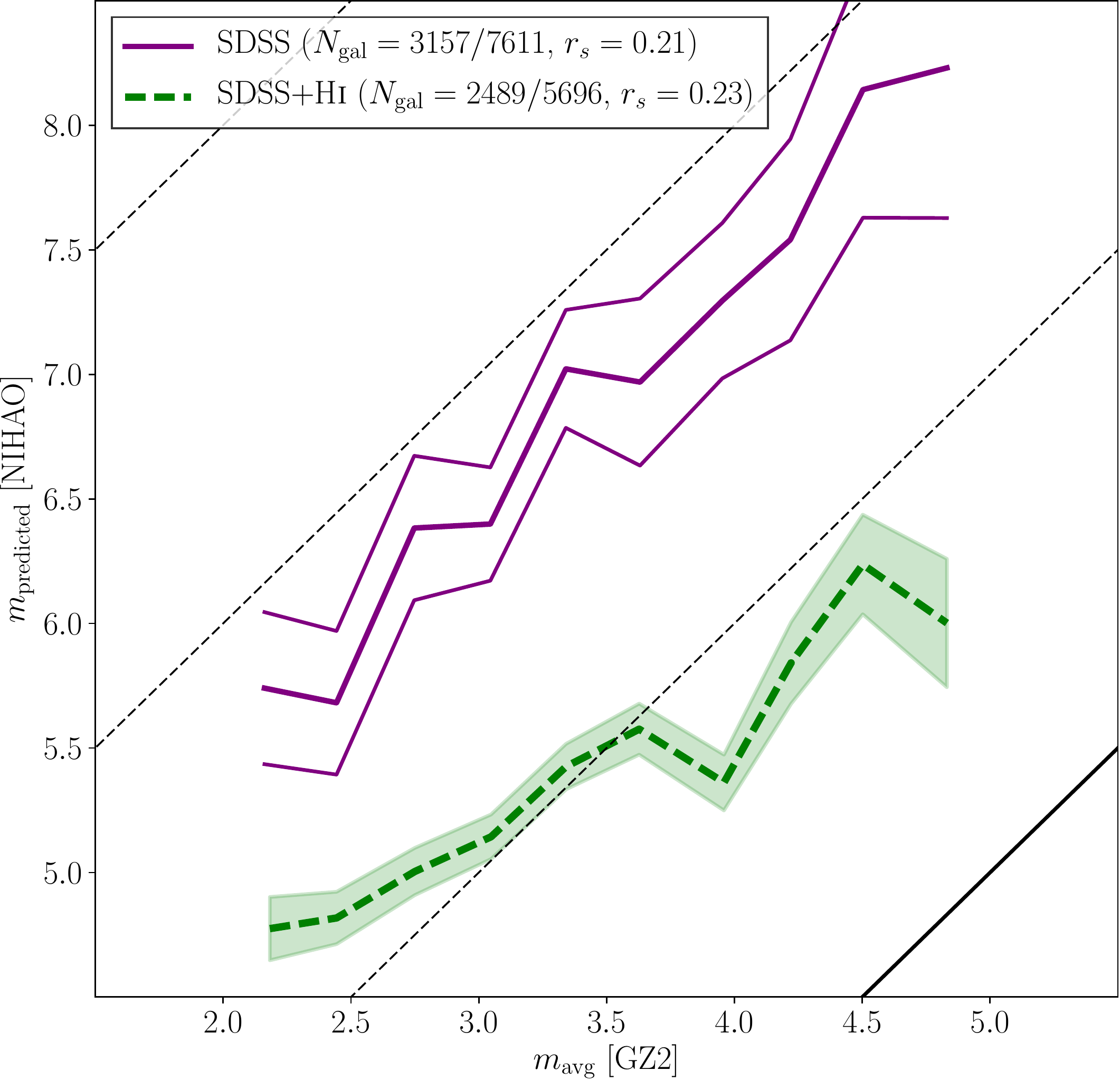}
\caption{Observed spiral arm number from GZ2 ($m_\mathrm{avg}$) vs. predicted spiral arm number for the SDSS and \sdsshi{} samples with haloes contracted or extended according to Eq.~\ref{eq:nihao_contraction}. The thick black line indicates the expected one-to-one correlation, and the dashed black lines show the same correlation offset by two spiral arms. Using this prescription for the dark matter halo, a key prediction from swing amplification emerges, suggesting that a fraction of the galaxies with realistic $\alpha$ values are swing amplified.}
\label{fig:m_vs_m_nihao}
\end{figure}

Given that these galaxies lie so close to the line defined by the NIHAO simulation, we test whether this relationship between galaxy size and flattened profile can produce arm numbers that one would expect if swing amplification was responsible for spiral arms. For all of the galaxies with physical $\alpha$ values, we contract or expand the haloes to match the prescription of Eq.~\ref{eq:nihao_contraction}. From these haloes, we again calculate the expected spiral arm number for the SDSS and the \sdsshi{} samples. The plot comparing predicted vs. observed spiral arm number is shown in Fig.~\ref{fig:m_vs_m_nihao}. From the resulting plot, we see a remarkable correlation: galaxies where one would expect to see more spiral arms do indeed have more arms. If the model were to work perfectly, then these galaxies would lie on the one-to-one line shown by the solid black line of Fig.~\ref{fig:m_vs_m_nihao}. Instead, the SDSS sample lies $\sim$3 spiral arm numbers too high, but with a strong correlation of $r_s=0.21$. If we include the \hi{} component in the disc, the model predicts spiral arm numbers more accurately, with the systematic offset reduced to $\sim$2 spiral arms, with a similar strength of correlation with $r_s=0.23$. The swing amplified model can predict a key observable, albeit with a systematic offset. The source or sources of this offset are discussed in more detail in Sec.~\ref{sec:discussion}.

\section{Discussion}
\label{sec:discussion}

By drawing on a number of observational measurements and models for dark matter haloes, we have investigated whether predictions of swing amplification theory can predict morphological characteristics of spiral arms in galaxies. Neither universal cusped or cored haloes can predict the spiral arm numbers or pitch angles in galaxies accurately. However, by invoking a halo which is contracted or expanded by an amount dependent on the relative size of its baryonic content, there is a population of galaxies for which our predicted spiral arm numbers correlate strongly with those observed.

\subsection{Can the model produce realistic spiral arms?}
\label{sec:model_realism}

In the local Universe, the varieties of spiral structure and their relative fractions are well constrained. The majority of spirals tend to have grand design arms -- both infra-red and optical studies show that $\approx 60$ per cent of unbarred, low-redshift spirals with stellar mass $\log(\mathrm{M_\odot}) \gtrsim 10$ have two-arm or grand design spiral structure \citep{Elmegreen_82,Grosbol_04,Hart_16}. In Sec.~\ref{sec:arm_number_distributions}, we demonstrated a familiar problem with the simulations of swing amplified spiral galaxies. These models produced galaxies with too many spiral arms, with median arm number of $\sim$10 spiral arms for the SDSS sample. The inclusion of the \hi{} component to add to the galaxy disc mass does improve the situation somewhat, reducing the median spiral arm number to $\sim$6. This picture is still unsatisfactory in terms of describing the spiral arms in our galaxy sample. Although an extra gas component can reduce the spiral arm number, it still cannot reproduce the dominant two-arm spiral population we expect to see. The other complicating factor is the role that gas plays in the disc. The swing amplified quantities described in this paper are based on N-body simulations -- the discs consist of many stellar particles, and their self-gravity form spiral arms in galaxies. The role that a gas component will play is not fully understood. The inclusion of a gas component can help to stabilise the two-arm mode \citep{Bournaud_02} and make swing amplification more efficient \citep{Jog_92,Jog_93}. Gas has also been suggested as a requirement to cool the stellar system in order for it to be unstable to arm formation \citep{Sellwood_85}.

In order to produce a modal galaxy population which produces a reasonable number of unbarred two-arm modes, we require that the dark matter halo potential in the inner regions is significantly reduced. In Sec.~\ref{sec:arm_number_distributions} we showed that the swing amplification mechanism can produce a spiral population with more prominent lower order modes if the dark matter halo is cored to the extent that it is flat within $a_\mathrm{h}$. Both the SDSS and \sdsshi{} models produce spiral arms representative of those at low-redshift. This does, however, produce its own complications. Of greatest concern is how stable such low-order modes are. Currently, N-body simulations cannot produce long-lived $m=2$ spirals without quickly forming a central bar \citep{Athanassoula_02,Sellwood_11,Dobbs_14}. Additionally, these modes do not predict a correlation between predicted and observed spiral arm numbers. Rather, the dominant mode is usually driven down to $\sim$2 in almost all spiral galaxies: a cored dark matter halo cannot predict the range of spiral arm numbers observed in low-redshift galaxies.

In order to reproduce realistic spiral arms, we have found that a halo with some level of interpolation between a cored and cusped dark matter halo is required. In order for the model to match the observations, most galaxies need some level of halo expansion. Such a result should not be surprising, however -- recent work suggests that low-redshift disc galaxies require strongly cored inner profiles in order to fit rotation curves \citep{Cole_17,Katz_17}. 

Examining these required halo sizes leads one to the conclusion that there are two distinct populations of galaxies. Of all of the galaxies, only $\sim 40$ per cent can be modelled by swing amplification with any kind of dark matter halo. Remarkably, these galaxies show a strong correlation between the spiral arm numbers expected and those observed. This leads us to conclude that swing amplification does play a dominant role in generating spiral structure in around half of unbarred disc galaxies. The secondary population is discussed in more detail in Sec.~\ref{sec:bimodality}.

Although a correlation does exist, it is offset from the one-to-one line that one would expect, overestimating the number of spiral arms by approximately three. This may be due to how mass is assigned to the bulge and disc. We use photometric decompositions of \citet{Simard_11} and \citet{Mendel_14} to assign mass to the bulge and the disc. Such a model fits a classical bulge with $n=4$ and an exponential disc. This may cause a systematic for two reasons. Firstly, the photometric decomposition of galaxies may introduce a bias due to image resolution effects. The second issue is the pseudo vs. classical bulge argument -- the model we use assumes an inner classical spherical bulge; bulges instead may be pseudo-bulges, which may not have a spherical shape, and profile well-described by a spherical Hernquist profile \citep{Carollo_97,Gadotti_01,Kormendy_06,Fisher_08,Gadotti_09}. Studying bulges and discs in detail is beyond the scope of this paper. Another possibility is that the assumption that spiral arms are measured at $2 R_\mathrm{d}$ may not be valid -- if spiral arms were instead measured closer to the inner regions of galaxies, then this offset is negated. Unfortunately, the binary nature of visual morphological classifications, where arms either are or are not recorded, prevents further investigation of this point. Finally, there may be some spiral arms which are impossible to observe with visual morphology in the way presented in this paper. Of particular note is the case where the model predicts very high spiral arm numbers. In this case, the spiral arms may instead be wakelets which are difficult to observe visually; our observed arm number measurements may therefore be systematically low for these galaxies. Investigating which caveat, or which combination of caveats is responsible requires higher resolution imaging of galaxies than those used in this paper. Any study of this nature would be severely restricted in terms of sample size and completeness compared to the results we present in this paper.

Another potential source of systematic uncertainty is the model we use to contract/expand the dark matter halo. We use a recently-published prediction from a full-hydrodynamical code in NIHAO \citep{Wang_15} to compare our model to predictions. However, there are a number of parameters that go into such a code -- the expansion of dark matter cores is usually driven by gas outflows caused by feedback from stars and supernovae \citep{Read_05,Governato_10,Pontzen_12,Chan_15}. Small adjustments to the strength of this feedback would theoretically cause haloes to expand more if the other properties of galaxies were kept the same. However, testing these effects is not the purpose of this paper.

\subsubsection{A note on disc maximality}
\label{sec:disc_maximality}

In a number of simulations, it has been shown that the spiral arm number has a dependence on the disc fraction within 2.2 disc scale radii, 2.2$R_\mathrm{d}$, which takes the functional form $m \propto f_\mathrm{d}^{-1}$ \citep{Carlberg_85,Bottema_03,Fujii_11,Donghia_13}. We therefore expect galaxies with greater disc fractions to have fewer spiral arms. 

Maximal discs are usually defined as discs with $f_\mathrm{d}(2.2R_\mathrm{d})>0.7$. Disc maximality has been a subject of much debate, and often depends on the technique one uses to measure it \citep{Bosma_17}. Recent work based on velocity dispersion measurements of disc galaxies suggests that discs may be sub-maximal \citep{Bottema_93,Kregel_05,Bershady_11}. A recent study that should be directly comparable to our work is that of \citet{Martinsson_13}, where a set of low-redshift spiral galaxies were decomposed into bulge and disc components, meaning the disc contribution was directly measured with little or no bulge contamination. This study also found discs to be sub-maximal, with $f_\mathrm{d}(2.2R_\mathrm{d})= 0.31 \pm 0.07$. However, other recent studies instead suggest that discs may indeed be more maximal than these kinematic studies would suggest. Velocity dispersion-based techniques rely on  estimates of both the velocity dispersion and the disc scale height, which are often probed by different stellar populations \citep{Bosma_99,Aniyan_16,Bosma_17,Aniyan_18}. \citet{Aniyan_16} demonstrated that accounting for these systematic differences leads one to conclude that the Milky Way's disc is maximal, in agreement with measurements of the Milky Way's rotation curve from \citet{Bovy_13}.

\citet{Athanassoula_87} showed that for swing amplified spiral arms to exist in a sample of nearby spirals, then a maximal disc is required to match the observed spiral arm numbers. We can see from Fig.~\ref{fig:all_distributions} that cored inner profiles mean that discs are close to maximal in nature, particularly if one considers the H\textsc{i} component in the SDSS galaxies. The work presented in this paper supports this idea that spiral arms can be maintained via a swing amplified mechanism, if the inner profiles of galaxies are cored to the extent of, or perhaps even more so than the relation given by the NIHAO prediction of \citet{Dutton_16}. If we assume that all of the galaxies which have realistic dark matter haloes described in Sec.~\ref{sec:varying_halo} exist via the swing amplified mechanism, then we obtain $f_\mathrm{d} =$ \upperlower{0.48}{0.11}{0.17} (\upperlower{0.54}{0.11}{0.15} with the inclusion of H\textsc{i}; the value quoted is the median, and the bounds indicate the 16th and 84th percentiles). The predictions using the NIHAO simulations, which generally give too many spiral arms, give $f_\mathrm{d} =$ \upperlower{0.21}{0.11}{0.19} (\upperlower{0.25}{0.19}{0.08}). In order for our spirals to be swing amplified modes, our discs must be more maximal than the sub-maximal values measured in \citet{Martinsson_13}.

\subsubsection{A bimodality in the galaxy population}
\label{sec:bimodality}

Given the evidence listed above, most spiral galaxies may not exist as swing amplified modes. Approximately 60 per cent of the galaxies in our sample do not fit the expected characteristics from swing amplification theory. We demonstrated that this sub-population cannot exist with the model we use in this paper -- even if there is no massive dark matter component, the spiral arm numbers predicted are still greater than those observed. A likely scenario is that spiral arms can be triggered and exist via a multitude of mechanisms, and that the model presented in this paper is only applicable to a select sample of galaxies. 

One mechanism that can generate spiral structure is the presence of bars. However, in this paper, we explicitly control for this by removing any galaxies with even weak bars in the various samples using visual galaxy classifications. However, we cannot rule out the other often quoted mechanisms for driving spiral arms: density wave theory and tidal interactions. Density wave theory \citep{Lin_64} is a mechanism via which two-arm spiral patterns can emerge, while simulations also predict that galaxy-galaxy interactions can effectively trigger the formation of two-arm patterns \citep{Toomre_72,Oh_08,Dobbs_10}. A second population of spiral galaxies completely separate from the swing amplified spirals also goes a long way to explain the results of many of the simulations to date. As yet, there are no simulations that predict long-lived, stable two-arm patterns in simulations of isolated galaxy discs \citep{Dobbs_14}. That two-arm spirals form a secondary population triggered in another way would suggest that the models are correct in that they do not predict two-arm patterns, and that inclusion of other physical parameters will not affect this result in future simulations.

\subsection{Arm number and pitch angle as tracers of swing amplification}
\label{sec:m_vs_psi}

We have the option to test the modal mechanisms of spiral structure using either pitch angle, arm number or a combination of both. We note that we expect our measured pitch angles to be less certain than the measures of spiral arm number, given that they require an accurate measurement on each individual galaxy compared to simply counting arms. We therefore suggest that the spiral arm number is the best technique for testing and calibrating any models of spiral galaxies. The different effectivenesses of pitch angle and arm number as tracers of swing amplification appears to be due to what they probe in the model. From Eq.~\ref{eq:donghia_m}, we see that spiral arm number is a result of the relative sizes and masses of the components that make up galaxies. However, pitch angle probes something altogether more subtle. The shear in galaxy discs probes the gradients of the mass distribution inside galaxies. With the models employed in this paper, all galaxies tend to have flat rotation curves, consistent with observations of the overall galaxy population. Without direct measurements of the galaxy dynamics from accurate galaxy rotation curve data, one cannot model the subtle differences that lead to large variations in pitch angles.

The spiral arm pitch angle distributions of our spiral galaxy populations were compared in Sec.~\ref{sec:pitch_angle_distributions}. The galaxy model we use in this paper leads to the majority of spirals having arms centred around $\psi=24$\textdegree. As was the case for the spiral arm number, the pitch angle is a quantity that we can constrain from observations of local galaxies. We used two complementary datasets to test how well the swing amplified predictions match the observations of local spirals. The \sfg{} sample was measured directly from mid infra-red imaging by hand by professional astronomers; the SDSS pitch angles were instead measured automatically, a method we tested the reliability of in \citet{Hart_17b}. We see that both samples give distributions centred on $\sim19$\textdegree{} with spiral arms ranging from 5-40\textdegree. These are both similar to the pitch angle distributions measured in other samples, indicating that these pitch angle distributions seem to be characteristic of the total galaxy population \citep{Seigar_98b,Block_99,Seigar_05,Seigar_08}. We see that neither dark matter halo can produce the correct distribution of spiral arm pitch angles: the distributions are too loose. We interpret this as evidence that spiral arm pitch angle depends on a number of different properties, rather than simply the underlying mass distribution in a swing amplification regime. The fact that the spiral arms are tighter than those predicted is also of interest, given the predictions for how spiral arms should evolve over time. Generally, spiral arms produced in a swing amplified N-body regime will tighten as the galaxy rotates \citep{Perez-Villegas_12,Grand_13}. It could potentially be the case that new spiral arms will form at $\psi=24$\textdegree{}, and age to produce the tighter distribution we observe in our galaxies. For our sample, a scatter introduced by this effect resolves the differences between the distributions of observed and predicted spiral arm numbers. Accurately estimating the age of a spiral arm would, however, be very challenging. Given the above caveats, spiral arm number was a much better method for testing the predictions of swing amplification, and was thus employed for the rest of this paper.

\section{Conclusions}
\label{sec:conclusions}

In this paper, we have compared direct predictions from a model of swing amplified spiral arm characteristics to those observed. By drawing on a number of measured data, and well-defined scaling relations where these are unavailable, we model a sample of galaxies from both the SDSS and the \sfg{} surveys. We find that using a simple cored or cusped profile to model the dark matter cannot account for the majority of spiral arms -- cusped profiles predict too many arms, whereas cored profiles cannot predict the complete variation in spiral structure across the galaxy population. However, by including a dark matter profile with some level of expansion, as predicted by simulations due to halo expansion caused by feedback from star formation, a significant agreement emerges -- approximately half of galaxies have spiral arms consistent with the model we employ in this paper. These display a significant correlation between predicted and observed spiral arm number. The rest of the unbarred spiral population is unlikely to be dominated by a swing amplified arms, and are instead more likely to be due to tidal interactions or density waves.

\section*{Acknowledgements}

The data in this paper are the result of the efforts of the Galaxy Zoo 2 volunteers, without whom none of this work would be possible. Their efforts are individually acknowledged at \url{http://authors.galaxyzoo.org}.

The development of Galaxy Zoo was supported in part by the Alfred P. Sloan foundation and the Leverhulme Trust.

RH acknowledges a studentship from the Science and Technology Funding Council, and support from a Royal Astronomical Society grant. RH acknowledges helpful discussion with Elena D'Onghia regarding the models presented in this paper. We also thank the referee for their insights which improved this paper.

Plotting methods made use of \texttt{scikit-learn} \citep{scikit-learn} and \texttt{astroML} \citep{astroML}. This publication also made extensive use of the \texttt{scipy} Python module \citep{scipy}, \texttt{TOPCAT} \citep{topcat}, \texttt{Astropy}, a community-developed core Python package for Astronomy \citep{astropy} and the \texttt{Uncertainties} Python package \citep{uncertainties}. 

Funding for SDSS-III has been provided by the Alfred P. Sloan Foundation, the Participating Institutions, the National Science Foundation, and the U.S. Department of Energy Office of Science. The SDSS-III web site is http://www.sdss3.org/.

SDSS-III is managed by the Astrophysical Research Consortium for the Participating Institutions of the SDSS-III Collaboration including the University of Arizona, the Brazilian Participation Group, Brookhaven National Laboratory, Carnegie Mellon University, University of Florida, the French Participation Group, the German Participation Group, Harvard University, the Instituto de Astrofisica de Canarias, the Michigan State/Notre Dame/JINA Participation Group, Johns Hopkins University, Lawrence Berkeley National Laboratory, Max Planck Institute for Astrophysics, Max Planck Institute for Extraterrestrial Physics, New Mexico State University, New York University, Ohio State University, Pennsylvania State University, University of Portsmouth, Princeton University, the Spanish Participation Group, University of Tokyo, University of Utah, Vanderbilt University, University of Virginia, University of Washington, and Yale University.

We thank the many members of the ALFALFA team who have contributed to the acquisition and processing of the ALFALFA data set.




\bibliographystyle{mnras}
\bibliography{bibliography} 



\appendix

\section{The Burkert dark matter profile}
\label{sec:burkert_profile}
In Sec.~\ref{sec:dark_matter_haloes}, a Burkert dark matter halo was discussed to model the dark matter halo of spiral galaxies. The Burkert profile \citep{Burkert_95} is characterised by the following function: \begin{equation}
\label{eq:burkert_profile}
\rho = \rho_0 \frac{r_0^3}{(r+r_0)(r^2 + r_0^2)} ,
\end{equation} where $\rho_0$ is the central density of the dark matter halo, $r_0$ is the scale length and $r$ is the radius. We define the following quantity to make the equations appear a little simpler: \begin{equation}
\phi (r,r_0) = \ln \Big(\frac{r+r_0}{r_0} \Big)
+ \frac{1}{2} \ln \Big( \frac{r^2+r_0^2}{r_0^2} \Big) 
- \arctan \Big(\frac{r}{r_0} \Big) .
\end{equation} The mass enclosed within a sphere of radius $r$ is given by \begin{equation}
\label{eq:burkert_mass_enclosed}
\mathrm{M}(r) = 2 \pi \rho_0 r_0^3 \phi(r,r_0).
\end{equation} The central density, $\rho_0$, can be calculated from the mass of the halo at $r_{200}$, where $r_{200}$ is the radius where the halo density falls to 200 times the critical density of the Universe. By rearranging Eq.~\ref{eq:burkert_mass_enclosed}, the central density is \begin{equation}
\label{eq:rho0}
\rho_0 = \frac{\mathrm{M_{200}}}{2 \pi r_0^3 \phi(r_{200},r)} ,
\end{equation} where $\mathrm{M_{200}}$ is the halo mass at $r_{200}$. The angular frequency of the halo is \begin{equation}
\label{eq:omega2_burkert}
\Omega^2 (r) = 2 \pi G \rho_0 r_0^3 \frac{\phi(r,r_0)}{r^3} .
\end{equation} The spiral arm number is given by Eq.~1 of \citet{Donghia_15}: \begin{equation}
\label{eq:donghia_m_raw}
m = \frac{\kappa^2}{2 \pi G \Sigma} \frac{R}{X} , 
\end{equation} where $\Sigma$ is the surface density of the stellar disc and $X$ is a factor introduced in \citet{Toomre_81} which is most effective at $X=1.5-2$ \citep{Donghia_13}. $\kappa^2$ is given by \begin{equation}
\label{eq:kappa2}
\kappa^2 = r \frac{d \Omega^2}{dr} + 4 \Omega^2.
\end{equation} For the Burkert profile, this becomes \begin{equation}
\label{eq:kappa2_burkert}
\kappa^2 = \frac{2 \pi G \rho_0 r_0^3}{r^3} \Big[
\frac{r^2}{r^2+r_0^2} - \frac{r}{r_0(r^2/r_0^2+1)} + \frac{r}{r+r_0} + \phi(r,r_0) \Big] .
\end{equation} Putting together Eq.~\ref{eq:donghia_m_raw} and Eq.~\ref{eq:kappa2_burkert} yields the following relation for spiral arm number with respect to the dark matter halo: \begin{equation}
\label{eq:burkert_arm_number} 
m_\mathrm{h} = \frac{e^{2y}}{X} \frac{\pi \rho_0 r_0^3}{\mathrm{M_d}} \frac{y^2}{2} \Big[
\frac{r^2}{r^2+r_0^2} - \frac{r}{r_0(r^2/r_0^2+1)}+ \frac{r}{r+r_0}
 + \phi(r,r_0) \Big] .
\end{equation}

This now replaces the halo term in Eq.~\ref{eq:donghia_m} so that $m$ can be calculated for the Burkert dark matter halo.

\section{The use of one or two component fits}
\label{sec:fit_test}

\begin{figure}
\centering
\includegraphics[width=0.45\textwidth]{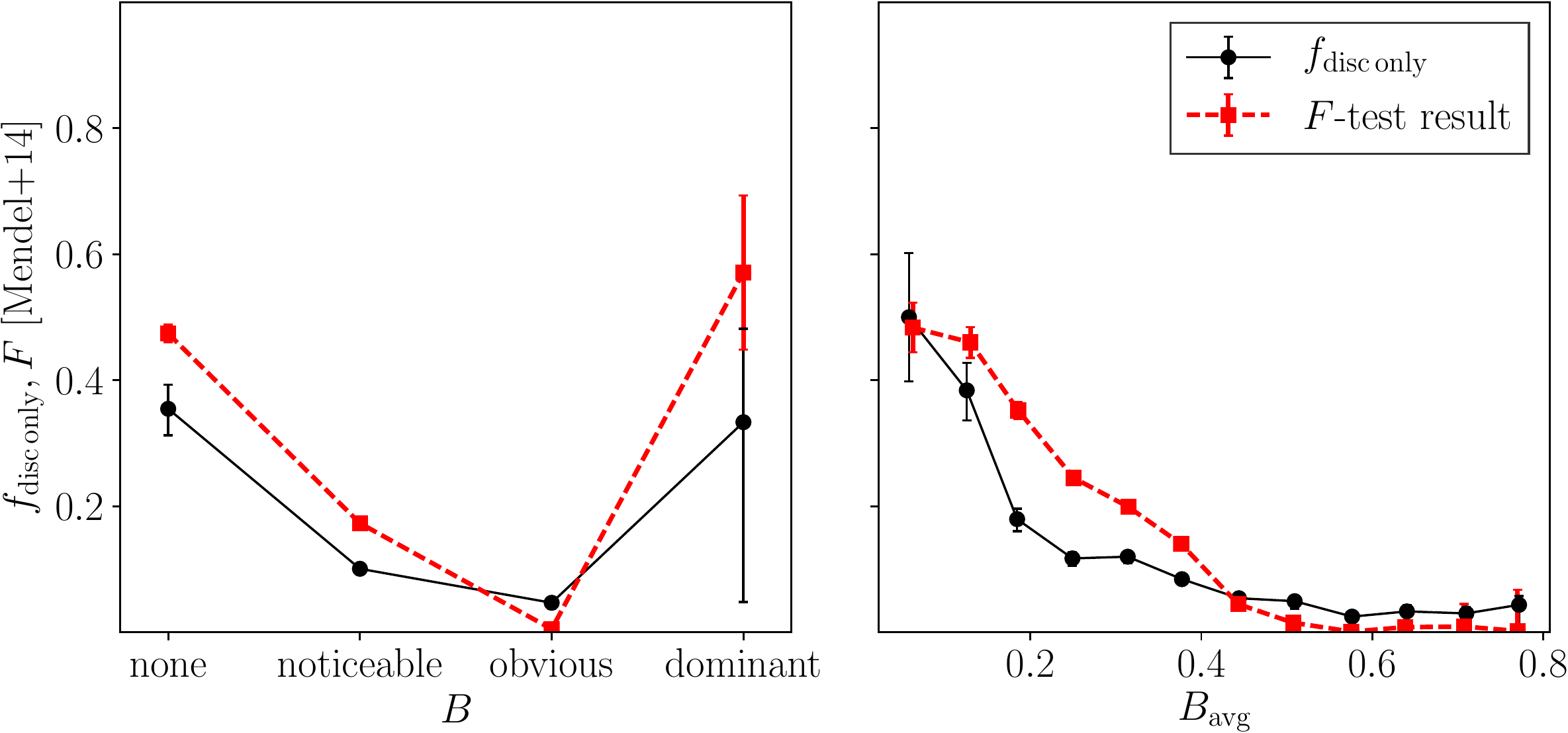}
\caption{(a) Modal value of response to the bulge prominence question in GZ2, $B$, vs. fraction of discs classified with only a disc (black circle points with a thin line) and median $F$-test value (red squares with a thicker dashed line). (b) As in (a), but using the continuous average bulge prominence statistic, $B_\mathrm{avg}$. We see a strong correlation that galaxies with less prominent bulges in GZ2 are more likely to be fit with a single disc component in \citet{Simard_11}.}
\label{fig:bulge_test}
\end{figure}

We use already derived bulge and disc mass estimates from \citet{Simard_11} and \citet{Mendel_14} as parameters in our galaxy model. The \citet{Mendel_14} catalogue usually fits galaxies with two components, but also includes single fit models. The catalogue provides a statistic, the $F$-test statistic, to determine which model is more appropriate. The paper also advises that this statistic is not perfect, and arguments from what is expected from the physical properties of galaxies should instead be used if possible. We do, however, want to avoid the fitting of a bulge+disc to a system where the galaxy has little or no bulge. The reliability of galaxy bulge size measurements have already been shown to correlate well with visually characterised bulge prominence statistic (Masters et al. in prep.). In order to check whether the $F$-test statistic can reliably identify bulge-less systems, we use the same visual statistic. We define the bulge prominence using the `is there any sign of a bulge?' question in GZ2. We define $B_\mathrm{avg}$ in the same way as Masters et al. in prep.:\begin{equation}
\label{eq:b_avg}
B_\mathrm{avg} = 0.0 p_\mathrm{no \, bulge} + 0.2 \cdot p_\mathrm{noticeable} + 0.8 \cdot p_\mathrm{obvious} + 1.0 \cdot p_\mathrm{dominant} ,
\end{equation} and the statistic $B$ which corresponds to which response to the bulge prominence question got the most votes. In Fig.~\ref{fig:bulge_test}, we check how both the median $F$-test statistic and the fraction of galaxies with $F<0.32$, $f_\mathrm{disc \, only}$, change with GZ2 bulge prominence. Here we see a strong correlation ($r_s(B_\mathrm{avg},f_\mathrm{disc \, only})=-0.41$) between the two statistics, meaning that galaxies with a higher probability of having no bulge from the GZ2 statistics are much more likely to require the single component model. We therefore use the $F$-test statistic to define whether we use a bulge+disc or disc only model for our SDSS galaxies. 

\bsp	
\label{lastpage}
\end{document}